\documentclass{article}
\usepackage{amsmath,amsfonts,color}

\def\C{\mathbb{C}}
\def\mm{\medskip\\}
\def\pas{\par\smallskip}
\def\lupm{L^{\scriptscriptstyle\uparrow}_{\scriptscriptstyle+-}}
\def\ludp{L^{\scriptscriptstyle\uparrow \downarrow}_{\scriptscriptstyle+}}
\def\ludpm{L^{\scriptscriptstyle\uparrow\downarrow}_{\scriptscriptstyle +-}}

\renewcommand{\theequation}{\thesection.\theequation}
\numberwithin{equation}{section}
    \title{On possible effects of the spinor structures in Quantum Physics}

    \author{Elena Ovsiyuk\footnote{Mosyr State Pedagogical University,
    Belarus, e.ovsiyuk@mail.ru},
 Olga Veko\footnote{Kalinkovichi Gymnasium, Belarus,vekoolga@mail.ru},
 Alexandru Oana\footnote{Transilvania University of Brasov, alexandru.oana@unitbv.ro},\\
            Mircea Neagu\footnote{Transilvania University of
            Brasov, mircea.neagu@unitbv.ro},
            Vladimir Balan\footnote{University Politehnica of Bucharest, Romania, vladimir.balan@upb.ro},
             Victor Red'kov\footnote{B.I. Stepanov Institute
of Physics, NAS of Belarus, redkov@dragon.bas-net.by}}
    \date{}
\begin{document}

\maketitle
\begin{abstract}The paper discusses the following topics: spinor coverings for the full Lorentz group,
    intrinsic parity of fermions, Majorana fermions, spinor structure of space models,
    two types of spacial spinors, parametrization of spinor spaces by curvilinear coordinates,
    manifestation of spinor space structure in classifying  solutions of the
    quantum-mechanical equations and in the matrix elements for  physical quantities.
\end{abstract}

\pas\noindent {\bf MSC2010}: 15A66, 78A25, 35Q60,
78A99.\pas\noindent {\bf Key-words}: spinors; tetrads;
Dirac--Schwinger quantization; Dirac equation; superposition
principle;
    polarized light; Majorana 4-spinors; Lorentz group; relativistic fermion parity.
%


\section*{General introduction}
In the literature [2--39], there exist three different terminological approaches,
    whose intrinsic essence is similar: a space-time spinor structure (Penrose,
    Rindler et al.); the Hopf  bundle and the Kustaanheimo-Stiefel bundles.\par
In Hopf's technique, one uses only complex 2-spinors ($\xi$)
    and their conjugates ($\xi^{*}$), instead of real-valued 4-vector (tensor) objects.
In the Kustaanheimo-Stiefel approach, there are used four real-valued coordinates --
    the real and imaginary parts of 2-spinor components.\par
The formalism developed in the present work exploits as well the possibilities given by spinors to
    construct 3-vectors; however, the emphasis is put on doubling the set of spacial
    points, so that we get an extended space model. In such a space, instead of the $2\pi$-rotation,
    there is considered the $4\pi$-rotation - which transfers the space into itself.
The procedure of extending the set of manifold points
    is achieved much easier, by using curvilinear coordinates.\par
Within the framework of applications of spinor theory to Relativistic and Non-relativistic
    Physics, Quantum Mechanics and Polarization Optics, we discuss several actual issues, as:
\begin{itemize}\setlength{\parskip}{-1mm}
\item the concept of spinor structure in space-time models;
\item exact linear representations for spinor coverings of the full Lorentz group;
\item internal space-time parity of a relativistic fermion;
\item Euclidean 3-spaces with opposite $P$-properties and two kinds of Cartan spacial spinors;
\item parametrization of Cartan's spacial spinors by curvilinear coordinates;
\item the role of spinor space structure in classification of solutions of the
    Klein-Fock-Gordon equation and the influence on the matrix elements related to physical quantities.
\end{itemize}
%
%
\section{ Spin covering for the full Lorentz group $\ludpm$
    and the concept of relativistic fermion parity }
To treat the problem of fermion parity, we will use 4-spinors instead of 2-spinors.
    Additional motivation for this approach is that among 4-spinors there exist real-valued ones --
    the so-called Majorana 4-spinors; moreover, in this way we will be able to describe
    discrete symmetries by linear transformations\footnote{We will mainly consider
    only the problem of accurate description of the single-valued representations
    of four different spinor groups, each of them covering the full Lorentz group
    $L^{\uparrow \downarrow}_{+-}$, including $P$ and $T$-reflections.}.\par
The obtained results will provide the grounds for a new discussion of the old
    fermion parity problem of investigating possible linear single-valued
    representations of spinor coverings of the extended Lorentz group. It is shown that
    in the frame of this theory, $P$-parity and $T$-parity for a fermion do not exist
    as separate concepts; instead of these, only some unified concept of ($PT$)-parity
    can be described in a group-theoretical language.\pas
We attach to the proper orthochronous Lorentz matrices
    \begin{equation}\begin{array}{l}L^{\;\;b}_{a}(k, k^{*})=\bar{\delta }^{b}_{c} (- \delta^{b}_{c}
    k^{n} k^{*}_{n}+k_{c} k^{b*}+k^{*}_{c} k^{b}+i \epsilon^{\;\;bmn}_{c} k_{m}
        k^{*}_{n}) ,\mm L(k, k^{*})=L(-k, -k^{*})\end{array}\end{equation}
two linear operations
    $${ P: }\;\; L^{(P)b}_{a}=+ \bar{\delta}^{b}_{a} \;;\qquad +T:\;\;
        L^{(T)b}_{a}=- \bar{\delta }^{b}_{a} ,$$
where $\bar{\delta}^{b}_{a}=\mbox{diag}(+1,-1,-1,-1)$, of which one readily produces the full Lorentz group $\ludpm$. The commutation rules between
    $L^{\;\;b}_{a}(k, k^{*})$ and the discrete elements $P,T$ are
    \begin{eqnarray} \bar{\delta}^{b}_{a} \;L^{\;\;c}_{b}(k, k^{*})=L^{\;\;b}_{a}(\bar{k}^{*}\; ,
    \bar{k})\;\bar{\delta }^{c}_{b} \; . \label{23.1c}\end{eqnarray}
The group $\ludpm$ has four types of vector representations:
    \begin{eqnarray}T^{\;\;b}_{a} (L)={f(L) }\; L^{\;\;b}_{a} \; , \;\; L \in
        \ludpm\; , \label{23.2a}\end{eqnarray}
namely
    \begin{equation}\begin{array}{ll}{f_{1}(L) }=1 \; ,&{f_{2}(L)}=\det (L) \;,\mm
    {f_{3} (L)} =\mbox{sgn} \; (L^{\;\;0}_{0})\; ,&{ f_{4}(L) }= \det
        (L)\; \mbox{sgn} \; (L^{\;0}_{0})\; . \end{array}\label{23.2c}\end{equation}
which have the explicit form
    \begin{eqnarray}\left. \begin{array}{llll} {1}:\qquad&T_{1}(L) =\;\; L &\qquad
    {2}:\qquad&T_{2}(L)=\;\;L\\&T_{1}(P)=+ P&& T_{2}(P)=- P\\
   &T_{1}(T)=+ T&& T_{2}(T)=- T\\ [2mm]{3}:&T_{3}(L)=\;\; L&\qquad
    {4}:&T_{4}(L)=\;\; L\\&T_{3}(P)=+ P&& T_{4}(P)=- P\\
   &T_{3}(T)=- T&& T_{4}(T)=+ T. \end{array}\right. \label{23.3}\end{eqnarray}
It should be emphasized that the above-described extension of the group
    $L^{\;\; b}_{a}(k,k^{*})$ by adding the two discrete operations $P$ and $T$ is not an
    extension of the spinor group $SL(2,\C)$: actually this is just an expansion
    of the orthogonal group $L^{\uparrow }_{+}$. From the spinor point of view, the
    operations $P$ and $T$ are transformations which act on the space of 2-rank spinors, and
    {\em not} on the space of 1-rank spinors. Evidently, a more comprehensive study of
    $P,T$-symmetry can be done in the framework of first-rank spinors, when one extends
    the covering group $SL(2,\C)$ by adding spinor discrete operations.\pas
Now we can start solving this task. A covering group for the total Lorentz group can be
    constructed by adding two specific $4\times 4$-matrices to the known set
    of 4-spinor transformations of the group $SL(2,\C)$,
    \begin{eqnarray}S(k, \bar{k}^{*})=\left( \begin{array}{cc} B(k)&0\\ 0&
    B(\bar{k}^{*}) \end{array}\right) \in\widetilde{SL}(2,\C).\label{23.4}\end{eqnarray}
Those two new matrices are to be taken from the following
    \begin{eqnarray}M=\left( \begin{array}{cc} 0&I\\
    I&0 \end{array}\right) \; , \; \; M'=i M \; ,\qquad N
    =\left( \begin{array}{cc} 0&-i I\\+i I&0 \end{array}\right) \; , \; \;
    'N=i N. \label{23.5}\end{eqnarray}
Having added any two elements of the four ones, we provide the full extension of the
    group $\widetilde{SL}(2,\C)$, by means of two new operations only.
    Also, we note that since the group $L(2,\C)$ contains $-I$,
    the extension of the group by any two elements of $\{-M , -M',-N, -'N\}$, leads to the same result.
    However, if one takes any other \index{phase factor} phase factor, different from
    $+1,-1,+i,-i$ for $M , M', N, 'N$, then this will result in substantially new extended groups.\pas
The multiplication table for these four discrete elements is
    \begin{eqnarray}\left. \begin{array}{ccccc}&M&M '&N&'N\\ [4mm]
    M \;\; &\left( \begin{array}{cc} \;\;\;\;I&\;\;\;\;0\\\;\;\;\;0&\;\;\;\;I
    \end{array}\right) &\left( \begin{array}{rr}
    \;\;\;iI&\;\;\;\; 0\\ \;\;\;\; 0&\;\;\; iI \end{array}\right)
    &\left( \begin{array}{rr} +iI&\;\;0\\ \;\;0&-iI\end{array}\right)&   \left( \begin{array}{rr} \;-I&\;\;0\\ \;\;0&\;+I \end{array}\right)\\[4mm]
    M' \;\; &\left( \begin{array}{rr} +iI&\;\; 0\\ \;\; 0&+iI
    \end{array}\right) &\left( \begin{array}{rr} \;\;-I&\;\;\; 0
   \\ \;\;\; 0&\;-I \end{array}\right) &\left(
    \begin{array}{rr} \;-I&\;\;\; 0\\ \;\;\;0&\;+I \end{array}
   \right) &\left( \begin{array}{rr} -iI &\;\; 0\\ \;\;0&+iI
    \end{array}\right)\\ [4mm] N\;\; &\left(\begin{array}{cc} -iI&\;\;0\\ \;\;0
   &+iI \end{array}\right)&\left( \begin{array}{rr} \;+I&\;\;\;0\\ \;\;\;0&\;-I
    \end{array}\right) &\left( \begin{array}{rr} \;+I&\;\;\;0\\
    \;\;\;0&\;+I \end{array}\right) &\left( \begin{array}{rr}
    +iI&\;\;0\\ \;\;0&+iI \end{array}\right)\\ [4mm] 'N
     \;\; &\left( \begin{array}{cc} \;+I &\;\;\; 0\\
    \;\;\;0&\;-I \end{array}\right) &\left( \begin{array}{rr}
    +iI&\;\;0\\ \;\;0&-iI \end{array}\right) &\left (\begin{array}{rr}
    +iI&\;\;0\\ \;\;0&+iI \end{array}\right) &\left(\begin{array}{rr}
    \;-I&\;\;\; 0\\ \;\;\;0&\;-I \end{array}\right)\end{array}\right.\end{eqnarray}
Hence we obtain six covering groups,
    \begin{eqnarray}\left. \begin{array}{llll} G_{M} &=\{\; S(k,\bar{k}^{*})
    \uplus M \uplus M' \; \} \; ,&\qquad
    G_{N} &=\{\; S(k,\bar{k}^{*}) \uplus N \uplus \; 'N \;\} \; ,\mm
    G' &=\{ \; S(k,\bar{k}^{*}) \uplus M' \uplus N \; \} \;,&\qquad
    'G &=\{ \; S(k,\bar{k}^{*}) \uplus 'N \uplus M \; \} \; ,\mm
    G &=\{ \; S(k,\bar{k}^{*}) \uplus M \uplus N \; \} \;,
   &\qquad 'G' &=\{ \; S(k,\bar{k}^{*}) \uplus M' \uplus 'N \;\} \;,
    \end{array}\right. \label{23.7}\end{eqnarray}
with the corresponding multiplication tables
    \begin{eqnarray}\left. \begin{array}{rrrr}
    {G_{M}}:&M^2=+ I \; ,\;\;&M^{'2}=- I \; ,&\;\; M M' =(M') M \; ;\\
    {G_{N}}:&N^2=+ I \; ,\;\;&'N^2=- I \; ,&\;\; N ('N)=('N) N \; ;\\
    G':&M^{'2}=-I \; ,\;\;&N^2=+ I \; ,&\;\; (M') N=- N (M') \; ;\\
    'G:&('N)^2=- I\; ,\;\;&M^2=+ I \; ,&\;\; ('N) M=- M ('N) \; ;\\
    G:&M^2=+ I\; ,\;\;&N^2=+ I \; ,&\;\; M N=- N M \; ;\\
    'G':&(M')^2=-I \; ,\;\;&N^{'2}=- I \; ,&\;\; (M')
    ('N)=- ('N)(M')\;,\end{array}\right . \label{23.8} \end{eqnarray}
and
    \begin{eqnarray}F\; S(k,\bar{k}^{*})=S(\bar{k}^{*},k) \; F\; , \;\; F \in
    \{ M , M', N,'N \} \; . \label{23.9}\end{eqnarray}
One can notice that the multiplication lows for the groups $G_{M}$ and $G_{N}$ happen
    to coincide; the same happens for $G'$ and $'G$. This implies that the groups $G_{M}$
    and $G_{N}$ (and respectively $G' $ and $'G$) represent the same abstract group. Indeed,
    it is readily verified that $G_{M}$ and $G_{N}$ (and, also, $G'$ and $'G$), can be
    transformed into each other by a similarity transformation:
    \begin{eqnarray}G_{N}=A \; G_{M} \; A^{-1} \; : \;\; \;
    A \; S(k, \bar{k}^{*})=S(\bar{k}^{*},k) \; A \; , \nonumber \label{23.10a}\\
     A\; M\; A^{-1}=+ N \;,\; A \;M'\; A^{-1}=+ 'N \; ,\nonumber\\
    A=\mbox{const}\cdot\left( \begin{array}{cc}-i I&0\\ 0 &+I\end{array}\right) \; ; \label{23.10b}\\
    'G=A \;G'\; A^{-1}\; : \;\; A\;S(k, \bar{k}^{*})=S(\bar{k}^{*},k)\; A\; ,\nonumber\\
    \nonumber A\; M' \; A^{-1}=+ 'N \; ,\; A\; N \; A^{-1}=- M\; , \;\\
    A=\mbox{const}\cdot\left( \begin{array}{cc}-i I&0\\ 0 &+I\end{array}\right) \; .\end{eqnarray}
In other words, we define here only four different covering groups. Since in literature
    all the six variants are discussed, we shall accordingly trace all of them.
%
%
\subsection*{1.2. Representations of the extended spinor groups}
We shall construct now the exact linear representations of the groups
    $G_{M}, \;G_{N},\; G', \; 'G $, $\; G , \; 'G'$. It suffices to consider
    in detail only one group; for convenience, let this be $G_{M}$. Its
    multiplication table is
    \begin{eqnarray}M^2=- I \; ,\; M^{'2}=- I \; , \; M \; M'=M\; M\; ,\nonumber \label{24.1}\mm
    F\; S(k, \bar{k}^{*})=S(\bar{k}^{*}, k)\; F \; , \;\; (\;\; F =M\; , M'\;\;) \; ,\nonumber\mm
    (k_{1},\bar{k}^{*}_{1}) (k_{2}, \bar{k}^{*}_{2})=(<k_{1},k_{2}>,\;
        <\bar{k}^{*}_{1} ,\bar{k}^{*}_{2} >) \;.\end{eqnarray}
where the symbol $< \;,\; >$ stands for the known multiplication rule in the group $SL(2,\C)$:
    \begin{eqnarray}<k_{1}, k_{2}>=(k^{0}_{1} k^{0}_{2}+\vec{k}_{1} \vec{k}_{2}; \;
    \vec{k}_{1} k^{0}_{2}+k^{0}_{2} \vec{k}_{1}+i [ \vec{k}_{1}\vec{k}_{2} ]) \; .\end{eqnarray}
Let us look for the solution of the problem of constructing the simplest irreducible representations
    of the spinor groups as mappings of the form
       \begin{eqnarray}{ T(g)=f(g)\; g} \; ,\qquad g \in G_{M} \; ,\qquad f(g_{1})
    \cdot f(g_{2})=f(g_{1} \cdot g_{2}) \label{24.2a}\end{eqnarray}
where $f(g)$ is a numerical function on the group $G_{M}$. Substitution (\ref{24.2a})
    into (\ref{24.1}) yields
    \begin{eqnarray}[ f(M) ]^2= f(I) \; ,\qquad [ f(M') ]^2= f(-I) \; ,\qquad
    f(S(k,\bar{k}^{*}))=f(S(\bar{k}^{*},k))\; ,\nonumber\\
    f(S(k_{1}, \vec{k}^{*}_{1}))\; f(S(k_2), \vec{k}^{*}_{2}))=
    f (S (<k_{1}, k_{2}> \; , \; <\vec{k}^{*}_{1} \; , \;\vec{k}^{*}_{2} >))\; .\end{eqnarray}
There exist four different such functions $f_{i}$, described by:
    \begin{eqnarray}\left. \begin{array}{ccccc}
    G_{M}&f_{1}(g)=& f_{2}(g)=& f_{3}(g)=& f_{4}(g)=\\[2mm]
    S(k,\bar{k}^{*}) &+1 &+1 &+1 &+1\\M &+1&- 1 &+1&- 1\\
    M' &+1&- 1&- 1 &+1 ,\end{array}\right. \label{24.2b}\end{eqnarray}
which provide four representations $T_{i}(g)$ of the group $G_{M}$.\pas
In the same manner, one can construct the analogous representation $T_{i}(g) $ of the
    remaining five groups. All these are described by the following table
    \begin{eqnarray}\left.
    \begin{array}{cccccc}&g&T_{1}(g)&T_{2}(g) &T_{3}(g)&T_{4}(g)\\ [4mm]
   &S(k,\bar{k}^{*})&S(k,\bar{k}^{*}) &S(k,\bar{k}^{*})&S(k,\bar{k}^{*})&S(k,\bar{k}^{*})\\
    G_{M}&M &+M&- M &+M&- M\\& M' &+M'&- M'&- M' &+M'\\[2mm]
    G_{N}&N &+N&- N &+N&- N\\& 'N&+'N&-'N&-'N&+'N\\[2mm]
    G'&M' &+M'&- M' &+M '&- M'\\& N &+N&- N&- N &+N\\[2mm]
    'G&'N&+' N&-'N&+' N&-'N\\& M &+M&- M&- M &+M\\[2mm]
    G&M &+M&- M &+M&- M\\& N &+N&- N &+N&- N\\[2mm]
    'G'&M'&+M'&- M' &+M'&- M'\\& 'N&+'N&-' N&-' N&+' N
    \end{array}\right. \label{24.3}\end{eqnarray}
For each of these groups, one can ask whether the four representations
    $T_{i}(g)$ are equivalent, or not. With the help of the relations
    \begin{eqnarray}F=const\left(\begin{array}{cc}-I&0\\ 0 &+I \end{array}\right)
    \;,\;\; F \; S(k,\bar{k}^{*}) \; F^{-1}=S(k,\bar{k}^{*}) \; ,\nonumber \label{24.4a}\\
    F \; M\; F^{-1}=- M \; , \; F \; M'\; F^{-1}=-M'\; F\; , \;\nonumber\\
    N \; F^{-1}=- N \;, \; F\; 'N \;F^{-1} =- 'N, \;\end{eqnarray}
it is easily follows that the type $T_{2}(g)$ is equivalent to the type $T_{1}(g)$, as well,
    $T_{4}(g)$ is equivalent to $T_{3}(g)$:
    \begin{eqnarray} T_{2}(g)=F\; T_{1}(g) \;F^{-1} \;,\qquad T_{4}(g)=F \; T_{3}(g)\; F^{-1}
    \; . \label{24.4b}\end{eqnarray}
Summarizing, we have got to the following: for each of the six groups, only two non-equivalent
    representations $g\;\rightarrow \; T(g) =f(g)\; g$ are possible:
    \begin{eqnarray} T_{1}(g) \sim T_{2}(g) \; ,\qquad T_{3}(g) \sim T_{4}(g) \; .\end{eqnarray}
Evidently, this result does not depend on the explicit realization of the discrete spinor
    transformations.\pas
The above study of the exact linear representations of the extended spinor groups leads to
    a new concept of a space-time intrinsic parity of a fermion. In group-theoretical terms
    $P$-parity and $T$-parity do not have any sense, instead only their joint characteristic,
    that might be called $(PT)$-parity, can be defined in the group-theoretic framework.
%
%
\subsection*{1.3. Representations of the coverings for partly extended groups
    $\lupm$ and $\ludp$}
Now we are going to consider the problem of linear representations of the spinor groups
    that cover the partly extended Lorentz groups $\lupm$ and $\ludp$ (improper
    orthochronous and proper non-orthochronous, respectively). Such groups can
    be constructed by adding any matrix from $\{M , M', N , 'N\}$.\pas
The case of the orthogonal group $\lupm$ leads to
    \begin{eqnarray}&&T_{1}=T_{3} \;; \;\; L \Longrightarrow L=(\mbox{sgn} \;
    L_{0}^{\;\;0}) \; L \; \; , \label{25.1a}\mm
    &&T_{2}=T_{4}\;; \;\; L \Longrightarrow L=(\det L) L=(\det L)
    (\mbox{sgn} \; L_{0}^{\;\;0})\; L \;, \label{25.1b}\end{eqnarray}
and the case of the group $\ludp$ looks as
    \begin{eqnarray}&&T_{1}=T_{4}\; ; \;\; L \Longrightarrow L=(\det L)
    (\mbox{sgn} \; L_{0}^{\;\;0}) \; L \; , \label{25.2a}\mm
    &&T_{2}=T_{3} \; \;; \;\; \; L\Longrightarrow \; L=(\det L) \;
    L=(sgn \;L_{0}^{\;\;0}) \; L \; . \label{25.2b}\end{eqnarray}
With the use of one additional discrete operation, one can determine four extended spinor groups:
    \begin{eqnarray}\widetilde{SL}(2,\C)_{M}=\{\; S(k,\bar{k}^{*}) \; \oplus \; M \}\;\; \;
    \mbox{and so on } . \label{25.3}\end{eqnarray}
We conclude that the extended groups $\widetilde{SL}(2,\C)_{M}\; ,\; \widetilde{SL}(2,\C)_{N}$ turn
    out to be isomorphic. Analogously, $\widetilde{SL}(2,\C)_{M'}$ is isomorphic to
    $\widetilde{SL}(2,\C)_{'N}$. Each of them covers both $\ludp$ and $\lupm$:
    \begin{eqnarray}\widetilde{SL}(2,\C)_{M} \; \sim \; \widetilde{SL}(2,\C)_{N}\;\; , \;\;
    \widetilde{SL}(2,\C)_{M'} \; \sim \; \widetilde{SL}(2,\C)_{'N} \; .\end{eqnarray}
Now, we shall list the simplest representations of these groups. The obtained result is as
    follows: all the representations $T_{i}(g)$ from above, while confining them to
    sub-groups $SL(2,\C)_{M(N)}$ and $SL(2,\C)_{M',('N)}$, lead to representations
    changing into each other by a similarity transformation. In other words, in fact
    there exists only one representation of these \index{partly extended spinor groups}
    partly extended spinor groups. This may be understood as the impossibility to determine
    any group-theoretical parity concept ($P$ or $T$) within the limits of partly extended
    spinor groups.
%
%
\subsection*{1.4. On reducing spinor groups to a real form}
Till now we have considered all the spinor groups $ G_{M} \sim G_{N}\; , \; \;
    G' \sim 'G \; , \; \; G \; , \; \; 'G' \; $ as possible group covering candidates to
    for the full Lorentz group $\ludpm$. It is desirable to formulate
    some extra arguments in order to choose only {\em one} spinor group as a natural (physical) covering.\pas
Note that in the bispinor space a special basis can be found using
    the bispinor wave function
    \begin{eqnarray}\Phi_{M}(x)=\varphi (x)+i \xi (x) \;,\end{eqnarray}
which transforms under the action of the group $SL(2,\C)$ by means of real $(4\times4)$-matrices.
    Therefore, the real 4-spinors $\varphi(x)$ and $\xi (x)$, constituents of the complex-valued
    $\Phi_{M}(x)$, transform as independent irreducible 4-dimensional spinor representations.
    In physical context of real \index{Majorana} Majorana fermions, this reads as
    a group-theoretical permission to exist. But these arguments have been based
    only on continuous $SL(2,\C)$-transformations, while the idea is to extend them
    on discrete operations too. So we must find the answer to the question of which
    of the extended spinor groups of matrices can be reduced to real-valued forms.
    With this goal in mind, we write down the bispinor matrix a the form that does not depend on the
    randomly chosen basis\footnote{We employed above the Weyl basis.}:
    \begin{eqnarray}S={1 \over 2} (k_{0}+k^{*}_{0})+{1 \over 2} (k_{0}-k^{*}_{0})
    \gamma^{5} +(k_{1}+k^{*}_{1}) \sigma^{01}+(k_{1}-k^{*}_{1}) i \sigma^{23} +\nonumber\\
    (k_{2}+k^{*}_{2}) \sigma^{02}+(k_{2}-k^{*}_{2}) i \sigma^{31}+(k_{3}+k^{*}_{3})
    \sigma^{03}+(k_{3}-k^{*}_{3})i \sigma^{13}.\end{eqnarray}
Any Majorana basis satisfies the relations
    \begin{eqnarray}(\gamma^{a}_{M})^{*}=- \gamma^{a}_{M} \; , \;\; (\gamma^{5}_{M})^{*}=
    -\gamma^{5}_{M} \; , \;\; (\sigma^{ab}_{M})^{*}=\sigma^{ab}_{M} \;{\Longrightarrow S^{*}
   =S}\;. \label{26.5}\end{eqnarray}
It remains to write down all the used discrete (matrix) operations $M , M' N,'N$
    in terms of Dirac matrices:
    \begin{eqnarray}M=+ \gamma^{0} \; ,\qquad M'=+ i\; \gamma^{0} \;,\qquad
    N=+ i \;\gamma^{5} \; \gamma^{0} \; ,\qquad 'N=- \gamma^{5} \; \gamma^{0} \;.
    \label{26.4}\end{eqnarray}
In Majorana frames, the group (continuous and discrete)  operations obey the following properties
    \begin{eqnarray}S^{*}=S\; , \; M^{*}=- M \; , \; (M')^{*}=+ M' \; , \;
    N^{*}=- N \; , \; ('N)^{*}=+ 'N . \label{26.6}\end{eqnarray}
Thus, the six spinor groups behave under complex conjugation as indicated below
    \begin{eqnarray}\hspace{-5mm}\left. \begin{array}{cccccc} G_{M}&G_{N}&G' &'G&G&'G'\\ [1mm]
    S^{*}=S&S^{*}=S&S^{*}=S&S^{*}=S&S^{*}=S&S^{*}=S\\
    M^{*}=- M&N^{*}=-N&M'^{*}= +M'&'N^{*}=+'N&M^{*} =-M &{M'^{*}=+M }\\
    (M')^{*}=M'&'N^{*}=+'N&N^{*}=-N&M^{*}=-M&N^{*} =-N&{'M^{*} ='M}
    \end{array}\right.\end{eqnarray}
Only the group $'G'$ can be reduced to a real-valued form, and only this group allows
    real-valued spinor representations, namely the Majorana fermions\footnote{This variant coincides with
    the known in the literature Racah group.}.\par
%
%
\subsection*{1.5. Conclusion to Section 1}
    The problem of fermion parity is considered on the base of investigating possible single-valued
    representations of spinor coverings of the extended Lorentz group. It is shown that in the
    frame of this theory, there do not exist -- as separate concepts -- $P$-parity and $T$-parity for
    a fermion; instead of this, only some unified concept of \index{$(PT)$-parity} $(PT)$-parity
    can be determined in group-theoretical terms. Apparently, physics with spinor group
    significantly differs from the one based on the orthogonal group $\ludpm$, and only experiment
    can decide on this problem. It is needless to say that this task cannot be solved without a thorough
    theoretical analysis of possible experimental verifications, in both orthogonal and spinor approaches.
%
%
\section{Geometry of 3-spaces with spinor structure}
Our approach to examine the spinor structure of 3-space is based on the
    concept of spacial spinor, defined through taking the "square root"
    of a real-valued 3-vector. Two sorts of spacial spinors, according to the $P$-orientation
    of an initial 3-space, are introduced: proper-vector or pseudo-vector ones. These spinors,
    $\eta$ and $\xi$, turn out to be different functions of Cartesian coordinates. To
    have a spinor space model, one has to use a doubling vector space $ \{ \; (x_{1},x_{2},x_{3})\;
    \otimes \;\; (x_{1},x_{2},x_{3})' \;\; \} $. The information which is reachable here in the
    first place concerns non-relativistic physics in the frames of ideas on spinor space structure.\par
Spinor functions are in one-to-one correspondence with coordinates $x_{i}\oplus x_{i}'$, with the whole axis
    $$(0,0,x_{3}) \oplus (0,0,x_{3})'$$
removed; they exhibit an exponential discontinuity. Due to this reason, we shall consider
    the properties of spinor fields $\xi (x_{i}\oplus x_{i}') $ and $\eta(x_{i}\oplus x_{i}')$
    in terms of continuity with respect to geometrical directions in the neighborhood of every point.
    This points out the possible fruitful geometrization within the Finslerian framework.\pas
We shall further examine two sorts of spacial spinors, with the
use of: cylindrical parabolic, spherical and
    parabolic coordinates. Transition from vector to spinor models is achieved by doubling the
    parameterizing domain $G(y_{1},y_{2},y_{3}) \; \Longrightarrow \tilde{G}(y_{1},y_{2},y_{3})$
    with new identification rules on the boundaries. The differential equations satisfied by spacial spinor
    fields have been explicitly constructed. The use of curvilinear coordinates makes it easier
    to extend the formalism to curved (pseudo-Riemannian) models.
%
%
\subsection*{2.1. Two sorts of spacial spinors}
We will start with the well-known Cartan's classification of 2-spinors with respect to the
    spinor $P$-reflection:
    \begin{eqnarray}\widetilde{SU}(2)=\left\{\left. g \in SU(2) \oplus J=\left(\begin{array}{cc} i&0\\
    0&i \end{array}\right)\;\right|\mbox{ det}\; g=+1, \; \mbox{det} \; J=-1\right \},\label{1.1}\end{eqnarray}
which provides 2-component spinors of two sorts $T_{A}$:
    \begin{eqnarray}T_{1}: \; T_{1}(g)=g, \; T_{1}(J)=+ J \; ,\qquad T_{2}: \;
    T_{2}(g)=g, \; T_{2}(J)=- J \; . \label{1.2}\end{eqnarray}
There exist two ways to construct 3-vectors (complex-valued, in general) in terms of these 2-spinors:
    \begin{eqnarray}&&1.\;\;(\xi \otimes \xi^{*})= a \;+\; a_{j} \; \sigma^{j}\; \; , \;\;
    a=\sqrt{ a_{j} \; a_{j}} \; , \mbox{ pseudo-vector} \; ;\\
    &&2.\;\; (\eta \otimes \eta)=\;(c_{j}
    \;+\; i \; b_{j}) \;\sigma^{j}\; \;, \mbox{ vector} \; .\label{1.3}\end{eqnarray}
According to the way of taking the square root of the three real numbers -- components of a 3-vector
    $(x_{i})$, one obtains two different spacial spinors
    \begin{eqnarray}\xi \; \Longleftrightarrow \; a_{j} \; ,\qquad \eta\;
    \Longleftrightarrow \; c_{j} \;\; \mbox{or} \;\; (b_{j}) \; . \label{1.4}\end{eqnarray}
%
%
\subsection*{2.2. The pseudo-vector space $\Pi_{3}$ and the spacial spinor $\xi $}
This spinor model is based on the mapping
    \begin{eqnarray} \Pi_{3} =(a_{1},a_{2},a_{3}) \oplus(a_{1},a_{2},a_{3})'
    \; \; \Longrightarrow\;\;  \xi:
     \nonumber\\
    \xi=\left (\begin{array}{c}\sqrt{a+a_{3} } e^{-i\gamma/2}\\
    \sqrt{a-a_{3}} e^{+i\gamma /2} \end{array}\right) , \;\;
    e^{i\gamma }={ a_{1}+i a_{2} \over \sqrt{ a^2_{1}+a^2_{2} } } \; . \end{eqnarray}
It should be noted that in describing $\Pi^{+}_{0}$ and $\Pi^{-}_{0}$ there arise peculiarities:
    at the whole axis $a_{3}$, the relations contain the ambiguity $(0+i0)/0$ (and the expressions
    for $\xi$ will contain a mute angle variable $\Gamma: \; \gamma\rightarrow \Gamma$)
    \begin{eqnarray}\Pi^{+}_{0}\;\;:\qquad \xi^{+}_{0}=\left (\begin{array}{c}
    \sqrt{+2a_{3}} \; e^{-i\Gamma /2}\\ 0\end{array}\right) \; ,\qquad \Pi^{-}_{0}\;\;:\qquad
    \xi^{-}_{0}=\left (\begin{array}{c}0\\ \sqrt{-2a_{3} }\; e^{+i\Gamma /2} \end{array}\right) \;,
    \nonumber\\ {e^{i\Gamma } }=\lim_{a_{1},a_{2}\rightarrow 0} {a_{1}+i a_{2} \over
    \sqrt{ a^2_{1}+a^2_{2} }}\; ,\qquad a_{3}= 0 , \; \xi=\left (\begin{array}{c}
    \sqrt{ a^2_{1}+a^2_{2} }\; e^{-i\gamma /2}\\[2mm]
    \sqrt{a^2_{1}+a^2_{2} } \; e^{+i\gamma /2}\end{array}\right) \; .\end{eqnarray}
%
%
\subsection*{2.3. The proper vector space $E_{3}$ and the spacial spinor $\eta$}
This type of spacial spinor is based on the map
    \begin{eqnarray}(\eta \otimes \eta)=\;(c_{j}+i\; {b_{j}}) \sigma^{j}\;\; .\end{eqnarray}
The vector $\vec{b}$ covers the upper half-space $E^{+}_{3}$ twice; the spinor $\eta^{+}$ is given by
    \begin{eqnarray}\eta^{+}=\left (\begin{array}{c}
    \sqrt{b-(b^2_{1}+b^2_{2})^{1/2} } \; e^{-i\gamma /2}\\[2mm]
    \sqrt{b-(b^2_{1}+b^2_{2})^{1/2} } \; e^{+i\gamma /2}\end{array}\right) \; ,\qquad
    e^{i\gamma }={ b_{1}+ i b_{2} \over \sqrt{ b^2_{1}+b^2_{2}}} \;.\end{eqnarray}
The vector $\vec{b}$ covers a down half-space $E^{-}_{3}$ twice; the spinor $\eta^{-}$ is
    \begin{eqnarray}\eta^{-}=\left (\begin{array}{ll} \sqrt{b-(b^2_{1}+b^2_{2})^{1/2} }&
    \left [ -\sqrt{{b_{1}+ib_{2} \over (b^2_{1}+b^2_{2})^1/2}} \;\right ]^{*}\\[3mm]
    \sqrt{b+(b^2_{1}+b^2_{2})^{1/2} } &\left [ +\sqrt{ {b_{1}+ib_{2} \over (b^2_{1}+
    b^2_{2})^1/2}} \;\right ] \end{array}\right) .\end{eqnarray}
The spinor field $\eta$ is continuous at  the plane $b_{3}=0$:
    \begin{eqnarray}\eta^{+\cap -}=\left (\begin{array}{c} 0\\ \sqrt{2 (b_{1}+i \; b_{2}) }
    \end{array}\right).\end{eqnarray}
%
%
\subsection*{2.4. The spacial spinor $\xi_{a_{3}} (a_{1}+ia_{2})$ and Cauchy-Riemann analiticity}
It is natural to regard the components of spinor $\xi=\xi(a_{j})$ as complex-valued functions
    of $z=a_{1}+i a_{2}$ and of a real-valued function $a_{3}$:
    \begin{eqnarray}\xi^1= U^1+i V^1 \; , \;\;\xi^2=U^2+i V^2 \; . \end{eqnarray}
We obtain the modified Cauchy-Riemann relations
    \begin{eqnarray}{\partial U^1 \over \partial a_{1}}-{\partial V^1 \over
    \partial a_{2}}={1 \over 2} (a_{1} \cos {\gamma \over 2}+  a_{2} \sin {\gamma \over 2})
    \left [{1\over a \sqrt{a+a_{3}}}+{\sqrt{a+a_{3}} \over \rho^2}\right ] , \nonumber\\
    {\partial U^1 \over \partial a_{2}}+{\partial V^1 \over\partial a_{1}}={1 \over 2}
    (a_{2} \cos {\gamma \over 2}- a_{1} \sin {\gamma \over 2})\left [{1 \over a\sqrt{a+a_{3}}}+
    {\sqrt{a+a_{3}}\over\rho^2}\right ] ,\nonumber\\
    {\partial U^2 \over \partial a_{1}}-{\partial V^2 \over
    \partial a_{2}}={1 \over 2} (a_{1} \cos {\gamma \over 2}- a_{2} \sin {\gamma \over 2})
    \left [{1 \over a \sqrt{a-a_{3}}}-{\sqrt{a-a_{3}}\over\rho^2}\right ] ,\nonumber\\
    {\partial U^2 \over \partial a_{2}}+{\partial V^2 \over\partial a_{1}}= {1 \over 2}
    (a_{2} \cos {\gamma \over 2}+  a_{1} \sin {\gamma \over 2})
    \left [{1\over a \sqrt{a-a_{3}}}-{\sqrt{a-a_{3}} \over \rho^2}\right ] .\end{eqnarray}
For $\rho\;\rightarrow \;\infty $, the Cauchy-Riemann conditions still hold true.\pas
A special note should be given to the behavior of the spinor field $\xi^{i}$ along the half-plane
    $\{ a_{1} \ge 0 , a_{2}=0 \}^{a_{3}\neq 0}$.
    Here the spinor $\xi$ is not a single-valued function
    of spacial
    points of the pseudo-vector space $\Pi_{3}$
        without any explanation, as domain of a mapping at the beginning of the section.
    because its values depend on the direction from which one approaches those points.
%
%
\subsection*{2.5. Calculating $\nabla \xi$ and $\nabla_{\vec{n}} \; \xi$. The differential equation}
The spinor exhibit continuity properties. In order to point them out, let us calculate first the 2-gradient
    along an arbitrary direction
    \begin{eqnarray}\nabla \xi=({\partial \over \partial a_{1}} \;\xi\; , \;\;{\partial \over
    \partial a_{2}} \; \xi) ,\qquad \nabla_{\vec{n}}\; \xi=(\vec{n} \; \nabla \xi)\end{eqnarray}
in the neighborhood of an arbitrary point\footnote{we use the notation $ \vec{n} \vec{a}=n_{1}a_{1}+n_{2} a_{2} ,
    \vec{n} \times \vec{a}=n_{1} a_{2}-n_{2} a_{1} $.}:
    \begin{eqnarray}\nabla_{\vec{n}} \; \xi^1={1 \over 2}\;\left [\;{ (\vec{n}\; \vec{a}) \over
    a (a+a_{3})}\;+\; i \; {\vec{n} \times\vec{a} \over \rho^2}\;\right ] \; \xi^1\; , \;\;\nonumber\\
    \nabla_{\vec{n}} \;\xi^2={1 \over 2} \;\left [\; {(\vec{n}\;\vec{a}) \over a (a-a_{3})}\;-\; i\;
    {\vec{n} \times \vec{a}\over \rho^2}\;\right ] \xi^2\; .\;\;\;\end{eqnarray}
This can be considered as a basic equation that prescribes the explicit form of the spinor
    $\xi (\vec{a})$. This understanding seems to be interesting due to its mathematical potential.
%
%
\subsection*{2.6. Spinor $\eta$ and its differential equation}
Similar things can be done for other spacial spinors. In
particular, we derive the differential
    equations for a spacial spinor $\eta$
\begin{eqnarray}\nabla_{\vec{n}} \eta^1={1 \over 2 \rho }\left [-{1\over b} (\vec{n} \;\vec{b})+
    {i \over \rho } (\vec{n} \times\vec{b})\right ] \eta^1\; ,\nonumber\\
    \nabla_{\vec{n}} \eta^2={1 \over 2 \rho}\left [+{1\over b} (\vec{n} \;\vec{b})-{i \over \rho}
    (\vec{n} \times\vec{b})\right ] \eta^2\; .\end{eqnarray}
This can be considered as a basic equation which prescribes the explicit form of the spinor
    $\eta (\vec{b})$.
    %
%
\subsection*{2.7. Comparison of the models $\xi$ and $\eta$}
We shall further describe several qualitative distinctions between the spinor models
    $\xi$ and $\eta$. The two models of spinor spaces relative to the $P$-orientation rely on
    the different mappings $\xi$ and $\eta $ defined over the same extended domain $\tilde{G}(y_{i})$.
    The natural question is: how do these two maps relate? An answer can be found by comparing the
    derived formulas for $\xi $ and $\eta $.\pas
One answer emerges straightforward:
    \begin{eqnarray}\eta={1 \over \sqrt{2}} (\xi-i \; \sigma^2 \xi^{*})\;\qquad
    \mbox{or inverse}\qquad \xi={1 \over \sqrt{2}} \; (\eta \;-\; i \;\sigma^2 \eta^{*}) \; .\end{eqnarray}
An issue which needs special attention is the fact that complex conjugation enters these relations explicitly,
    fact which correlates to the change in orientation properties of the models.\pas
We have seen that the description of differently $P$-oriented geometries in terms of spinor
    fields $\eta$ and $\xi$ has made hardly noticeable the distinction between these two
    geometries - much more apparent and intuitively appreciable as connected with different types
    of spacial geometries which in vector description differ only in the alternative use of vectors
    and pseudo-vectors.
%
%
\subsection*{2.8. Spinors $\xi $ and $\eta $ in cylindrical parabolic coordinates}
This coordinate system in the vector $E_{3}$-space is defined by
    \begin{eqnarray}x_{1}={y^2_{1}-y^2_{2} \over 2 } \;\; ,\;\; x_{2}=y_{1}
    \; y_{2} \; \; , \;\; x_{3}=y_{3} \;\; ,\;\;\nonumber\\
    y_{2} \in [\; 0,+\infty\;) \;\; , \;\; y_{1},\; y_{3} \in (\;-\infty , \;+\infty \;) \; .\end{eqnarray}

\unitlength=0.5mm
\begin{picture}(120,70)(-100,0)
\special{em:linewidth 0.4pt} \linethickness{0.4pt}
\put(+10,+30){\vector(+1,0){100}} \put(+110,+25){$y_{1}$}
\put(+60,0){\vector(0,+1){70}} \put(+62,+70){$y_{2}$}
\put(+20,+30){\line(0,+1){30}} \put(+30,+30){\line(0,+1){30}}
\put(+40,+30){\line(0,+1){30}} \put(+50,+30){\line(0,+1){30}}
\put(+70,+30){\line(0,+1){30}} \put(+80,+30){\line(0,+1){30}}
\put(+90,+30){\line(0,+1){30}} \put(+100,+30){\line(0,+1){30}}
\put(+60,+30){\circle{2}} \put(+20,+30){\circle*{2}}
\put(+30,+30){\circle*{2}} \put(+40,+30){\circle*{2}}
\put(+50,+30) {\circle*{2}} \put(+70,+30){\circle*{2}}
\put(+80,+30) {\circle*{2}} \put(+90,+30){\circle*{2}}
\put(+100,+30){\circle*{2}}
\put(+60,+30){\oval(20,10)[b]} \put(+60,+30){\oval(40,20)[b]}
\put(+60,+30){\oval(60,30)[b]} \put(+60,+30){\oval(80,40)[b]}
\end{picture}

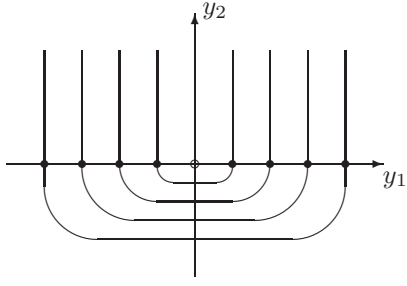
\begin{center}{\small\bf Fig. 1. Parabolic cylindrical coordinates}\end{center}
%


\noindent where the identified points on the boundary are connected by lines, and the domain
    $G(y_{1},y_{2})^{y_{3}}$ (at arbitrary $y_{3}$) ranging in the half-plane
    $(y_{1},y_{2})$ covers the whole vector plane $(x_{1},x_{2})^{x_{3}}$.\pas
The spinor $\xi $ of the pseudo-vector $\Pi_{3}$-model is given by
    \begin{eqnarray}\xi=\left (\begin{array}{c} \sqrt{(y^2_{3}+(y^2_{1}+
    y^2_{2})^2 / 4)^{1/2}+y_{3} } e^{-i\gamma /2}\\ [3mm]
    \sqrt{(y^2_{3}+(y^2_{1}+y^2_{2})^2 / 4)^1/2\;-y_{3}} e^{+i\gamma /2}
    \end{array}\right), e^{i\gamma /2}={y_{1}+i y_{2} \over \sqrt{ y^2_{1}+
    y^2_{2} }} ,\end{eqnarray}
the factor $e^{i\gamma /2}$ runs through the upper complex half-plane in the case of
    vector space. At the $x_{3}$-axis, we have:
    \begin{eqnarray}\xi^{+}_{0}=\sqrt{+2 y_{3}}\left (\begin{array}{c}
    e^{-i\Gamma /2}\\ 0 \end{array}\right) ,\qquad \xi^{-}_{0}
    = \sqrt{-2 y_{3}}\left (\begin{array}{c} 0\\ e^{+i\Gamma /2}\end{array}\right) \; ,\end{eqnarray}
For a proper vector model, the $\eta $-spinor looks as\footnote{The values $+$ and $-$ taken
    by the symbol $\sigma$ correspond to the $x_{3}>0$ and $x_{3}<0$ half-spaces, respectively.}
    \begin{eqnarray}\eta^{\sigma } (y)=\left (\begin{array}{c} \sqrt{\sqrt{y^2_{3}+
    (y^2_{1}+y^2_{2})^2/ 4}-{y^2_{1}+y^2_{2} \over 2}} \; \sigma e^{-i\gamma /2}\\[3mm]
    \sqrt{ \sqrt{y^2_{3}+(y^2_{1}+y^2_{2})^2 /4}+{y^2_{1}+y^2_{2} \over 2} } \;\;
    e^{-i\gamma /2}\end{array}\right) .\end{eqnarray}
We construct the extended (spinor) models $\tilde{E}_{3}$ and $\tilde{\Pi}_{3}$ by doubling the
    range of the $y_{2}$-variable:
    \begin{eqnarray}y_{2} \in [\; 0 , \;+\infty) \;\; \Longrightarrow \;\;
    y_{2} \in (- \infty , \;+\infty) \; .\end{eqnarray}
Then the above factor $e^{+i\gamma /2}$ will run through the full circle.

\vspace{5mm} \unitlength=0.6mm
\begin{picture}(120,70)(-72,0)
\special{em:linewidth 0.4pt} \linethickness{0.4pt}

\put(+10,+30){\vector(+1,0){100}} \put(+110,+25){$y_{1}$}
\put(+60,0){\vector(0,+1){70}} \put(+62,+70){$y_{2}$}
\put(+20,+30){\line(0,+1){30}} \put(+30,+30){\line(0,+1){30}}
\put(+40,+30){\line(0,+1){30}} \put(+50,+30){\line(0,+1){30}}
\put(+70,+30){\line(0,+1){30}} \put(+80,+30){\line(0,+1){30}}
\put(+90,+30){\line(0,+1){30}} \put(+100,+30){\line(0,+1){30}}
\put(+60,+30){\circle{2}}

\put(+20,+30){\line(0,-1){30}} \put(+30,+30){\line(0,-1){30}}
\put(+40,+30){\line(0,-1){30}} \put(+50,+30){\line(0,-1){30}}
\put(+70,+30){\line(0,-1){30}} \put(+80,+30){\line(0,-1){30}}
\put(+90,+30){\line(0,-1){30}} \put(+100,+30){\line(0,-1){30}}

\put(+60,+22){$ R^{exp.}$}
\end{picture}
\begin{center}{\small\bf Fig. 2. Space with spinor structure\footnote{
    The symbol $ R^{exp.}$ in Fig. 2 stands for the exponential discontinuity at
    all the axis $(0,0,x_{3})$.}}\end{center}
\noindent It is important to note the substantial change in the identification rules at the
    boundary set of $G(y_{1},y_{2},y_{3})$: for the extended domain
    $\tilde{G}(y_{1},y_{2},y_{3})$ one needs no special rules at all.
    Another issue needs to be emphasized: we have the same extended set
    $\tilde{G}(y_{1},y_{2},y_{3})$ for both spinor models $\xi(y)$ and $\eta (y)$.
    This means that only the providing of the set with doubling dimension and the using
    of identification rules, do not determine in full the whole geometry of the spinor spaces.
The specification of their $P$-orientation apparently requires additional information about
    this set. Evidently, $P$-orientation manifests itself in the explicitly different spinor
    functions $\xi(y)$ and $\eta (y)$. Moreover, a qualitative distinction between these
    spinor functions is revealed if one follows the orientation of a spinor
    $(\xi_{1},\xi_{2})$ and $(\eta_{1},\eta_{2})$, while going from the $x^{+}_{3}$ -- half-space
    to the $x^{-}_{3}$ -- half-space.\par
The differential equations for spacial spinors are
    \begin{eqnarray}\nabla_{\vec{\nu }} \; \xi^1={\xi^1\over 2}\left [ {\rho\over a (a+a_{3}) }
    (\vec{ n }\; \vec{y})+{i \over \rho }(\vec{n } \times \vec{y})\;\right ] ,\nonumber\\
    \nabla_{\vec{n }} \; \xi^2={\xi^2\over 2}\left [ {\rho\over a (a-a_{3}) } (\vec{n }\;
    \vec{y})- {i \over \rho } (\vec{n } \times \vec{y})\right ] ,\end{eqnarray}
and\footnote{We denote $(\vec{n } \; \vec{y})=n_{1} y_{1}+ n_{2} y_{2}$,
    $\vec{n } \times \vec{y}= n_{1} y_{2}-n_{2} y_{1}$.}
    \begin{eqnarray}\nabla_{\vec{n }} \; \eta^1={\eta^1\over 2}\left [-{\vec{n } \vec{y}
    \over b}+{i \over \rho } (\vec{n } \times\vec{y})\right ] ,\qquad
    \nabla_{\vec{n }} \; \eta^2={\eta^2\over 2}\left [{\vec{n }\; \vec{y}\over b}-{i
    \over \rho } (\vec{n } \times\vec{y})\right ] ,\end{eqnarray}
These equations have no peculiarities over the complex plane $y_{1}+i y_{2}$, excluding the origin $0+i0$.
%
%
\subsection*{2.9. The spinors $\xi$ and $\eta$ in parabolic coordinates}
We shall further describe the spinor approach relative to the well-known parabolic coordinates
    \begin{eqnarray}&&x_{1}=y_{1} y_{2} \; \cos y_{3}\; , \; \;x_{2}=y_{1} y_{2}\;
    \sin y_{3} \; ,\nonumber\mm&&x_{3}={y^2_{1}-y^2_{2} \over 2}\; , \;\;
    y_{1},y_{2} \in [ 0 ,+\infty)\; , \;\;y_{3} \in [ 0 , 2 \pi ]\end{eqnarray}
The spacial spinor $\eta$ of the proper vector model is given by
    \begin{eqnarray}\eta^{+}(y)={1 \over \sqrt{2}}\left (\begin{array}{c}
    (y_{1}-y_{2})\;\; e^{-iy_{3}/2}\\ (y_{1}+y_{2}) \;\; e^{+iy_{3}/2}\end{array}\right) , \;\;
    \eta^{-}(y)={1 \over \sqrt{2}}\left (\begin{array}{lr} (y_{2}-y_{1}) &(- e^{-iy_{3}/2})\\
    (y_{2}+y_{1})&e^{+iy_{3}/2}\end{array}\right) .\end{eqnarray}
As for the pseudo-vector model $\Pi_{3}$, we have
    \begin{eqnarray}\xi=\left (\begin{array}{c} y_{1} e^{-iy_{3}/2}\\ y_{2}
    e^{+iy^3/2} \end{array}\right) , \; \xi=\left (\begin{array}{c} N e^{-i\gamma /2}\\
    M e^{+i\gamma /2}\end{array}\right) ,\; y_{1}=N , y_{2}=M , y_{3}=\gamma\; .\end{eqnarray}
We double the above domain $G(y) \Longrightarrow \tilde{G}(y) (y_{3} \in [-2\pi ,+2\pi]$,

\vspace{4mm} \unitlength=0.5mm
\begin{picture}(110,100)(-75,0)
\special{em:linewidth 0.4pt} \linethickness{0.6pt}

\put(+60,+50){\vector(+1,0){70}} \put(+130,+45){$y_{2}$}
\put(+60,+50){\vector(0,+1){50}} \put(+53,+100){$y_{3}$}
\put(+60,+50){\vector(-1,-1){40}} \put(+25,+10){$y_{1}$}
\put(+45,+85){$+2 \pi $} \put(+55,+70){\circle*{2}}
\put(+55,+35){\circle*{2}} \put(+55,+70){\line(0,-1){35}}
\put(+55,+72){$R^{\pm}$} \put(+55,+30){$R^{\pm}$}
\put(+70,+80){\circle*{2}} \put(+70,+45){\circle*{2}}
\put(+70,+80){\line(0,-1){35}} \put(+72,+80) {$R^{\pm}$}

\put(+60,+85){\line(+1,0){60}} \put(+60,+85){\line(-1,-1){30}}
\put(+30,+55){\line(+3,+1){90}} \put(+120,+50){\line(0,+1){35}}
\put(+30,+20){\line(+3,+1){90}} \put(+30,+20){\line(0,+1){35}}

\put(+72,+45){$R^{\pm}$} \put(+10,+40){$R^{exp.}\;\Longrightarrow
$} \put(+110,+70){$\Longleftarrow \; R^{exp.}$}
\end{picture}
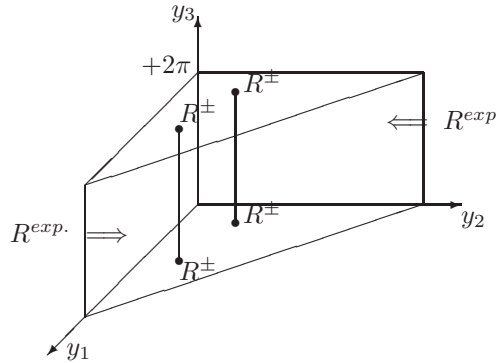
\begin{center}{\small\bf Fig. 3. Parabolic coordinates / vector space}\end{center}
\pagebreak
\vspace{-15mm}

\unitlength=0.5mm
\begin{picture}(140,100)(-80,+30)
\special{em:linewidth 0.4pt} \linethickness{0.4pt}

\put(+60,+50){\vector(+1,0){70}} \put(+130,+45){$y_{2}$}
\put(+60,0){\vector(0,+1){100}} \put(+53,+100){$y_{3}$}

\put(+60,+50){\vector(-1,-1){40}} \put(+25,+10){$y_{1}$}
\put(+45,+85){$ +2 \pi $} \put(+45,+15){$-2 \pi $}
\put(+55,+70){\circle*{2}} \put(+55,0){\circle*{2}}
\put(+55,+70){\line(0,-1){70}} \put(+70,+80){\circle*{2}}
\put(+70,+10){\circle*{2}} \put(+70,+80){\line(0,-1){70}}
\put(+10,+40){$R^{exp.}\;\Longrightarrow $}

\put(+60,+85){\line(+1,0){60}} \put(+60,+85){\line(-1,-1){30}}
\put(+30,+55){\line(+3,+1){90}} \put(+30,+20){\line(+3,+1){90}}

\put(+120,+15){\line(0,+1){70}} \put(+30,+20){\line(+3,+1){90}}
\put(+30,-15){\line(+3,+1){90}} \put(+30,-15){\line(0,+1){70}}

\put(+90,+60){$\Longleftarrow \; R^{exp.}$}
\put(+10,0){$R^{exp.}\; \Longrightarrow $}
\put(+90,+30){$\Longleftarrow \; R^{exp.}$}
\end{picture}

\vspace{20mm}
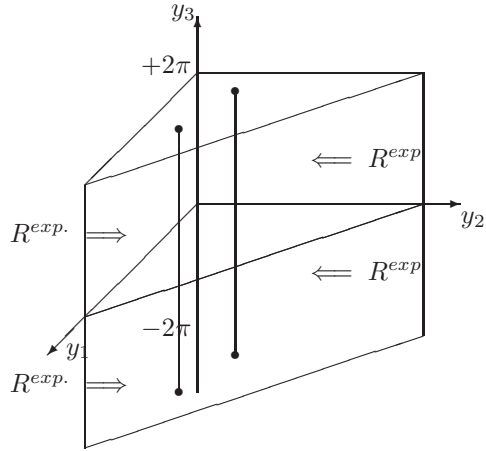
\begin{center}{\small\bf Fig. 4. Parabolic coordinates / spinor space}\end{center}
Instead of the domain $\tilde{G}(y)$ described below

\vspace{23mm} \unitlength=0.6mm

\begin{picture}(140,40)(-55,-20)
\special{em:linewidth 0.4pt} \linethickness{0.4pt}

\put(0,0){\vector(+1,0){40}} \put(+45,-5){$y_{1}$}
\put(0,0){\vector(0,+1){40}} \put(-5,+40){$y_{2}$}
\put(0,0){\line(1,1){40}} \put(0,+10){\line(1,1){30}}
\put(+10,0){\line(1,1){30}} \put(0,+20){\line(1,1){20}}
\put(+20,0){\line(1,1){20}}

\put(+55,0){$\oplus$}

\put(70,0){\vector(+1,0){85}} \put(130,0){\circle*{2}}
\put(90,0){\circle*{2}} \put(155,-8){$y_{3}$}
\put(110,0){\circle*{2}} \put(130,+5){$+2\pi$}
\put(90,5){$-2\pi$}
\end{picture}
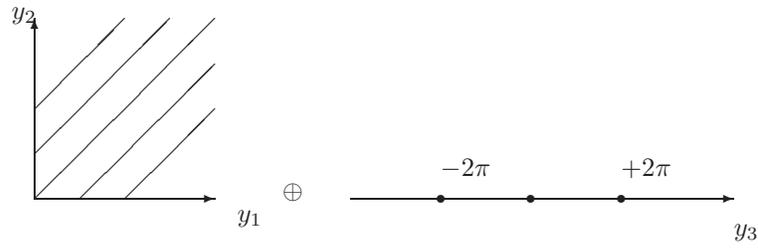
\begin{center}{\small\bf Fig. 5. Domain parameterizing the spinor space }\end{center}
one can use

\vspace{+20mm} \unitlength=0.5mm

\begin{picture}(140,50)(-100,-28)
\special{em:linewidth 0.4pt} \linethickness{0.4pt}

\put(-40,0){\vector(+1,0){80}} \put(+45,-5){$y_{1}$}
\put(0,-40){\vector(0,+1){80}} \put(-5,+40){$y_{2}$}
\put(0,0){\line(1,1){40}} \put(0,+10){\line(1,1){30}}
\put(+10,0){\line(1,1){30}} \put(0,+20){\line(1,1){20}}
\put(+20,0){\line(1,1){20}}

\put(0,0){\line(-1,-1){40}} \put(0,-10){\line(-1,-1){30}}
\put(-10,0){\line(-1,-1){30}} \put(0,-20){\line(-1,-1){20}}
\put(-20,0){\line(-1,-1){20}}

\put(+55,0){$\oplus$}

\put(70,0){\vector(+1,0){85}} \put(130,0){\circle*{2}}
\put(155,-8){$y_{3}$} \put(110,0){\circle*{2}}
\put(130,+5){$+2\pi$}
\end{picture}

\vspace{10mm}

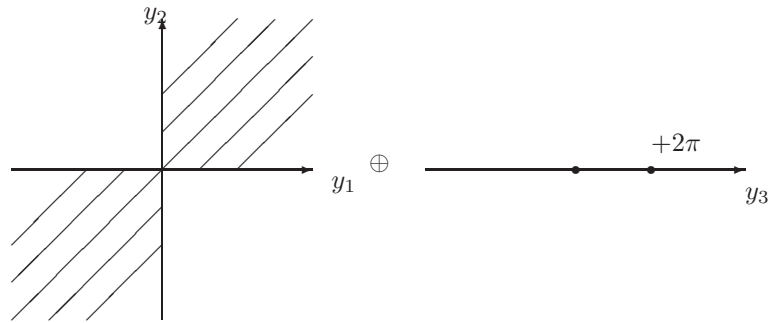
\begin{center}{\small\bf Fig. 6. Alternative domain to parameterize the spinor space }\end{center}

\vspace{6mm} Actually, various domains $\tilde{G}(y)$ are acceptable for the correct
    parametrization of spinor spaces, and one may choose any of them.
%
%
\subsection*{2.10. Spatial spinors in spherical coordinates}
We consider the system of spherical coordinates
    \begin{eqnarray}&&x_{1}= y_{1} \sin y_{2} \cos y_{3},\qquad x_{2}= y_{1} \sin
    y_{2} \sin y_{3} ,\qquad x_{3}= y_{1} \cos y_{2} \;,\nonumber\mm
    &&y_{1} \in [ 0,+\infty) \; ,\; y_{2} \in [ 0,+\pi ]\; ,\;y_{3} \in
    [0, +2\pi ]\; .\end{eqnarray}
A spinor $\eta(y)$ of the pseudo-vector model $\Pi_{3}$ is given by
    \begin{eqnarray}\xi=\left (\begin{array}{c}\sqrt{y_{1}(1+\cos y_{2}) } \; e^{-iy_{3}/2}\\
    \sqrt{y_{1}(1-\cos y_{2})} \; e^{+iy_{3}/2} \end{array}\right) .\end{eqnarray}
In turn, a spinor $\eta(y)$ of the proper vector model $E_{3}$ is defined according to
    \begin{eqnarray}\eta=\left (\begin{array}{lr}\sqrt{y_{1}(1-\sin y_{2})}&(\sigma e^{-iy_{3}/2}\\
    \sqrt{y_{1}(1+\sin y_{2})}&e^{+iy_{3}/2}\end{array}\right) \; ,\end{eqnarray}
The discontinuity properties of these spinors may be characterized by the diagram

\vspace{6mm}

\unitlength=0.7mm
\begin{picture}(100,50)(-80,+0)
\special{em:linewidth 0.4pt} \linethickness{0.4pt}

\put(0,+10){\vector(+1,0){70}} \put(+63,+5){$y_{3}$}
\put(+10,0){\vector(0,+1){50}} \put(+3,+48){$y_{2}$}
\put(+10,+30){\line(+1,0){40}} \put(+2,+30){$+\pi$}
\put(+50,+10){\line(0,+1){20}} \put(+48,+5){$+2\pi$}
\put(+10,+30){\line(+1,0){40}} \put(+30,+32){$R^{exp.}$}
\put(+53,+20){$\leftarrow R^{\pm 1}$} \put(+30,+3){$R^{exp.}$}
\put(-10,+20){$R^{\pm1}\rightarrow$ }
\end{picture}
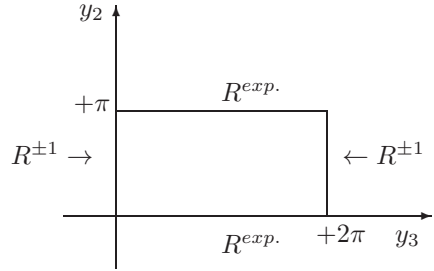
\begin{center}{\small\bf Fig. 7. Spherical coordinates in the vector space}\end{center}
Evidently, the transition to extended models can be performed by formal doubling
    the range of angle variable $y_{3}$\footnote{In the following we will use the
    more common notation $y_{1}=r , y_{2}=\theta , y_{3}=\phi)$}
    $$\tilde{G}(r,\theta ,\phi)=\{ \; r\in [ 0 ,+\infty) \;,\;\; \theta \in
    [ 0 ,+\pi ] , \phi \in [-2\pi ,+2\pi ]\; \} \; .$$
There are possible some alternative variants for the extended domain $\tilde{G}$, which can be used
    for covering spinor spaces. For instance, the most natural and symmetrical manner to do this,
    is to extend the range of radial variable:
    \begin{eqnarray}\tilde{G}'(r,\theta ,\phi)=\{ \; r \in (- \infty ,+\infty)\; ,\;\;
    \theta \in [ 0 ,+\pi ] , \phi \in [- \pi ,- \pi ] \; \} \;.\nonumber\\
    \xi (r,\theta ,\phi)=\left (\begin{array}{ll}
    \sqrt{1+\cos \theta }\;&(\; \sqrt{ r\; e^{i\phi } }\;)^{*}\\
    \sqrt{1-\cos \theta } \;& \; (\; \sqrt{ r \;e^{i\phi } } \;)\end{array}\right) \; .\end{eqnarray}

\unitlength=0.7mm
\begin{picture}(100,60)(-70,+0)
\special{em:linewidth 0.4pt} \linethickness{0.4pt}

\put(+0,+30){\vector(+1,0){100}} \put(+100,+25){$r$}
\put(+50,+0){\vector(0,+1){60}} \put(+43,+55){$\phi $}
\put(+10,+50){\line(+1,0){80}} \put(+90,+40){$A$}
\put(+10,+10){\line(+1,0){80}} \put(+90,+20){$B$}
\put(+51,+51){$+\pi $} \put(+38,+5){$-\pi $}
\end{picture}
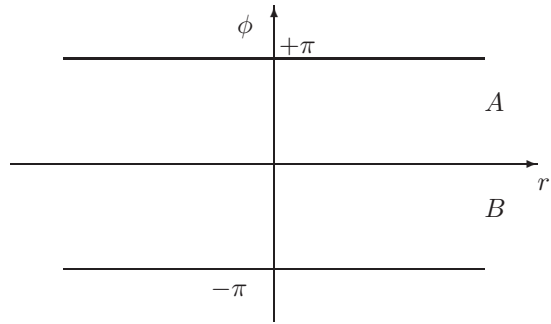
\begin{center}{\small\bf Fig. 8. Spherical coordinates in the spinor space}\end{center}
%
%
\subsection*{2.11. Conclusion to Section 2}
%
The results obtained for the 3-space with $(x,y,z)$ coordinates may be extended to
    Minkowski 4-space with coordinates $(t,x.y,z)$. Mathematically, this means to use
    the relativistic $SL(2,\C)$ spinors instead of non-relativistic $SU(2)$ spinors.
    The domains of curvilinear coordinates associated to spinor spaces can be used
    in order to examine possible quantum mechanical manifestation of the spinor structure,
    both in non-relativistic and relativistic theories. To this end, one should
    specially examine the analytical properties of the known solutions of the
    Schr\"{o}dinger and Dirac equations in various coordinates.
%
%
\section{Space with spinor structure and analytical properties of the \\solutions
    of the Klein--Fock--Gordon equation in parabolic cylindrical coordinates}
\subsection*{3.1. Parabolic cylindrical coordinates}
Let us start with the parabolic cylindrical coordinates
    \begin{eqnarray}x=(u^2-v^2)/2 \; ,\qquad y=u \; v\; ,\qquad z= z.\end{eqnarray}
In order to cover the vector space $(x,y,z)$, it suffices to make a choice out of the four possibilities:
    \begin{eqnarray}{v=+ \sqrt{-x+\sqrt{x^2+y^2} } \; ,\qquad u=
    \pm\sqrt{+x+\sqrt{x^2+y^2} }}\; ,\nonumber\\
    v=- \sqrt{-x+\sqrt{x^2+y^2} } \; ,\qquad u=\pm\sqrt{+x+\sqrt{x^2+y^2} }\; ,\nonumber\\
    v=\pm \sqrt{-x+\sqrt{x^2+y^2} } \; ,\qquad u=+\sqrt{+x+\sqrt{x^2+y^2} }\; ,\nonumber\\
    v=\pm \sqrt{-x+\sqrt{x^2+y^2} } \; ,\qquad u=-\sqrt{+x+\sqrt{x^2+y^2} }\; .\end{eqnarray}
For definiteness, let us use the first variant from the above: 
    \begin{eqnarray}v=+ \sqrt{-x+\sqrt{x^2+y^2} } \; ,\qquad u=\pm
    \sqrt{+x+\sqrt{x^2+y^2} }\; .\end{eqnarray}
which is illustrated in Figure 9.

\vspace{5mm}

\unitlength=0.45mm
\begin{picture}(100,50)(-160,0)
\special{em:linewidth 0.4pt} \linethickness{0.6pt}

\put(-50,0){\vector(+1,0){100}} \put(70,-5){$u$}
\put(0,-50){\vector(0,+1){100}} \put(-10, +45){$v$}

\put(+1,+1){\line(+1,+1){40}} \put(+21,+1){\line(+1,+1){40}}
\put(+21,+1){\line(+1,+1){40}} \put(-19,+1){\line(+1,+1){40}}
\put(-38,+1){\line(+1,+1){40}} \put(-58,+1){\line(+1,+1){40}}
\put(+41,+1){\line(+1,+1){40}}

\end{picture}

\vspace{25mm}
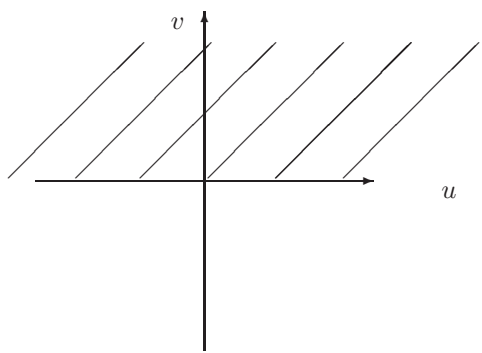
\begin{center}{\small\bf Fig. 9. The region $G(u,v)$ used to parameterize the vector model }\end{center}

The correspondence between the points $(x,y)$ and $(u,v)$ can be
    illustrated by the following formulas and Figure 10:
    \begin{eqnarray}&&u=k \; \cos \phi \; ,\qquad v=k \;\sin \phi \; ,\qquad
    \phi \in [\; 0 , \; \pi\; ] \; ;\nonumber\mm&&x=(k^2/2)\; \cos 2\phi \; ,\qquad
    y=(k^2/2) \; \sin 2\phi\; ,\qquad 2\phi \in [ 0 , \; 2 \pi ].\;\end{eqnarray}

\vspace{5mm}

\unitlength=0.5mm
\begin{picture}(100,50)(-80,0)
\special{em:linewidth 0.4pt} \linethickness{0.6pt}

\put(-50,0){\vector(+1,0){100}} \put(65,-5){$x$}
\put(0,-50){\vector(0,+1){100}} \put(-10, +45){$y$}

\put(-30,+25){$B_{1}$} \put(-30,-25){$B_{2}$}
\put(+25,+25){$A_{1}$} \put(+25,-25){$A_{2}$}

\put(+100,0){\vector(+1,0){100}} \put(215,-5){$u$}
\put(150,0){\vector(0,+1){50}} \put(140, +45){$v$}
\put(150,0){\line(+1,+1){40}} \put(150,0){\line(-1,+1){40}}

\put(180,+10){$A_{1}$} \put(110,+10){$A_{2}$}
\put(155,+20){$B_{1}$} \put(135,+20){$B_{2}$}

\put(-2,+20){$*$} \put(+165,+15){$*$}
\put(0,-20){\circle*{3}} \put(+135,+15){\circle*{3}}
\put(+30,0){\circle{6}} \put(+180,0){\oval(7,7)[t]}
\put(+120,0){\oval(7,7)[t]}

\end{picture}

\vspace{30mm}

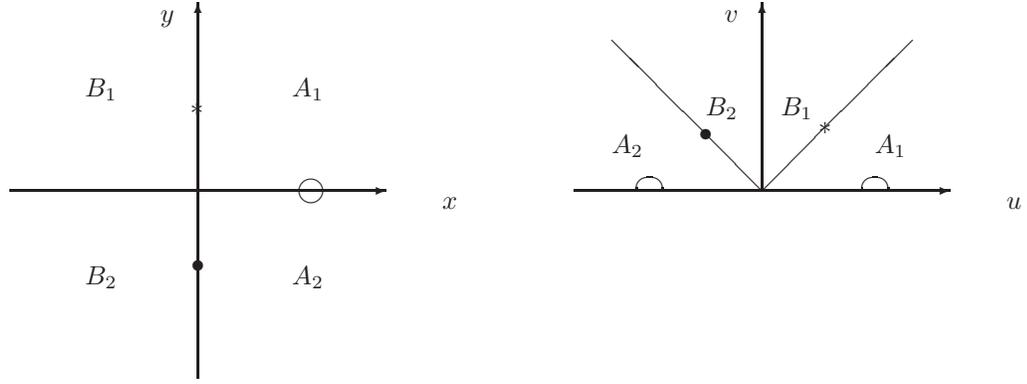
\begin{center}{\small\bf Fig. 10. The mapping $G(x,y) \Longrightarrow G(u,v)$;
    identification rules}\end{center}


When turning to the case of spinor space, we will see the complete symmetry
    between the coordinates $u$ and $v$; they relate to the Cartesian coordinates
    of the extended model $(x,y,z)\oplus (x',y',z')$ through the formulas
    \begin{eqnarray}v={\pm} \sqrt{-x+\sqrt{x^2+y^2} } \; ,\qquad u ={\pm}
    \sqrt{+x+\sqrt{x^2+y^2} }\;,\label{2.4}\end{eqnarray}
illustrated by Figure 11:

\vspace{5mm}

\unitlength=0.5mm
\begin{picture}(100,50)(-145,0)
\special{em:linewidth 0.4pt} \linethickness{0.6pt}

\put(-77,0){\vector(+1,0){150}} \put(85,-5){$u$}
\put(0,-50){\vector(0,+1){100}} \put(-15, +50){$v$}

\put(+1,+1){\line(+1,+1){40}} \put(+21,+1){\line(+1,+1){40}}
\put(+21,+1){\line(+1,+1){40}} \put(-19,+1){\line(+1,+1){40}}
\put(-38,+1){\line(+1,+1){40}} \put(-58,+1){\line(+1,+1){40}}
\put(+41,+1){\line(+1,+1){40}}

\put(+1,-1){\line(+1,-1){40}} \put(+21,-1){\line(+1,-1){40}}
\put(+21,-1){\line(+1,-1){40}} \put(-19,-1){\line(+1,-1){40}}
\put(-38,-1){\line(+1,-1){40}} \put(-58,-1){\line(+1,-1){40}}
\put(+41,-1){\line(+1,-1){40}}

\end{picture}

\vspace{28mm}

\begin{center}{\small\bf Fig. 11. $\tilde{G}(u,v)$ covering the spinor space}\end{center}

\noindent The metric of space-time in parabolic cylindrical coordinates has the form
    \begin{eqnarray}dS^2=c^2 dt^2-(u^2+ v^2) (du^2+ d v^2)-dz^2 \; .
    \end{eqnarray}
%
%
\subsection*{ 3.2. Solutions of the Klein--Fock--Gordon equation
    and functions on the parabolic cylinder}
Let us consider the KFG equation
    \begin{eqnarray}\left [ \;-{1 \over c^2} {\partial^2 \over \partial t^2}
    +{ \partial^2 \over \partial z^2}+{1\over u^2+v^2} \; \left({\partial^2 \over
    \partial u^2 }+ {\partial^2 \over \partial v^2 }\right)-{m^2c^2 \over \hbar^2}
    \right ] \Psi=0 .\label{3.1a}\end{eqnarray}
After separating the variables by the substitution
    $$ \Psi (t,u, v, \phi)=e^{-i\epsilon t / \hbar} \;e^{ipz/\hbar } \; U(u) \; V(v),$$
one gets
    \begin{eqnarray}\left [ \; {1 \over U } \; {d^2 U \over d u^2}+\left({\epsilon^2
    \over \hbar^2 c^2}-{ m^2 c^2 \over\hbar^2}-{p^2\over \hbar^2}\right) \; u^2 \;
    \right]+\;\;\;\;\nonumber\\
    +\left [ \; {1 \over V } \; {d^2 V \over d v^2}+\left({\epsilon^2\over \hbar^2 c^2}-
    {m^2c^2 \over\hbar^2}-{p^2\over \hbar^2}\right) \; v^2 \;\right ]=0 \; . \label{3.1b}\end{eqnarray}
In the following, we shall use the notation
    \begin{eqnarray}\lambda^2=\left({\epsilon^2 \over \hbar^2 c^2}-{ m^2
    c^2 \over \hbar^2}-{p^2\over \hbar^2}\right) \; ,\qquad [
    \lambda ]={1 \over \mbox{meter}} \; .\end{eqnarray}
By introducing two separation constants, $a$ and $b$ ($a+b=0$), we can derive from (\ref{3.1b})
    two distinct equations:
    \begin{eqnarray}{d^2 U \over d u^2}+(\; \lambda^2 \; u^2- {a }\;) \; U =0 \; ,\qquad
    {d^2 V \over d v^2}+(\; \lambda^2 \; v^2-{b}\;) \; V =0 \; .\label{3.2}\end{eqnarray}
The transition in equations (\ref{3.2}) to the canonical form is obtained by using dimensionless variables:
    \begin{eqnarray}\sqrt{2\lambda } \; u \;\;\rightarrow \;\; u \; , \; {a \over
    2\lambda } \;\;\rightarrow \;\; a \; ,\;\; \sqrt{2 \lambda } \;v \;\;\rightarrow
    \;\; v \; , \; {b \over 2 \lambda } \;\;\rightarrow \;\; b \; . \label{3.4}\end{eqnarray}
The equations (\ref{3.2}) will take the form:
    \begin{eqnarray}{d^2 U \over d u^2 } \;+\; \left(\; {u^2 \over 4}-a \;\right)\; U=0 \; ,\qquad
    {d^2 V \over d v^2 } \;+\; \left(\; {v^2\over 4} +a \;\right) \; V=0 \; . \label{3.5}\end{eqnarray}
The solutions of these similar equations can be found as series:
    \begin{equation}\begin{array}{ll}F(\xi)=&c_{0} \; +\; c_{1} \;\xi \; +\; c_{2}\; \xi^2 \;\mm
    &+ \; \sum_{k=1,2,...}\; c_{2k+1} \; \xi^{2k+1} \;+\;\sum_{k=1,2,...} \; c_{2k+2}
    \;\xi^{2k+2};\end{array}\label{3.6a}\end{equation}
we note that in (\ref{3.6a}) the terms of even and odd powers of $\xi$ are separated.\par
After tedious calculation, one derives two independent groups of recurrent relations:

\vspace{2mm}

\underline{for even powers}
    \begin{eqnarray}\left. \begin{array}{rl}
\xi^{0}:&\qquad 2 \; c_{2}-\alpha \; c_{0}=0 \; ,\mm
\xi^2:&\qquad c_{4} \; 4 \times 3+{c_{0} \over 4}-\alpha \; c_{2} =0 \; ,\mm
\xi^{4}:&\qquad c_{6}\; 6\times 5+{c_{2} \over 4}-\alpha \; c_{4} =0 \; ,\mm
n=3,4,... , \;\; \xi^{2n}:&\qquad c_{2n+2} (2n+2) (2n+1)+ {1 \over 4} \;
    c_{2n-2}-\alpha \; c_{2n} \;=0 \; ;\end{array}\right. \label{3.6b}\end{eqnarray}

\vspace{5mm}

\underline{for odd powers}
\begin{eqnarray}\left. \begin{array}{rl}
\xi^1:&\qquad c_{3} \; 3 \times 2-\alpha \; c_{1}=0 \; ,\mm
\xi^3:&\qquad c_{5} \; 5 \times 4+{c_{1} \over 4 }-\alpha \; c_{3}=0 \; ,\mm
n=3,4,... , \;\; \xi^{2n-1}:&\qquad c_{2n+1} (2n+1) (2n)+{1 \over 4} \;
c_{2n-3}-\alpha \; c_{2n-1}=0 \; .\end{array}\right. \label{3.6c}\end{eqnarray}

So one can construct two linearly independent solutions\pas

\underline{even}
    \begin{eqnarray}{F_{1}}(\xi^2)=1+a_{2}{\xi^2 \over 2!}+a_{4}
    {\xi^{4}\over 4!}+... ,\nonumber\mm
    a_{2}=\alpha \; ,\qquad a_{4}=\alpha^2 -{1 \over 2}\; ,
    \qquad c_{6}=\alpha^3-{7\over 2} \alpha \; ,\nonumber\mm
    n=3,4,...:\qquad a_{2n+2}=\alpha \; a_{2n}-{(2n)(2n-1)
    \over 4}\; a_{2n-2} \; ;\label{3.8a}\end{eqnarray}

\underline{odd}
\begin{eqnarray}F_{2} (\xi)=\xi+a_{3}\; {\xi^3 \over 3!}+a_{5}\;{ \xi^{5}
    \over 5!}+... \; ,\nonumber\\
    \nonumber a_{3}=\alpha \; ,\qquad a_{5}=\alpha^2-{3\over 2} \; ,\\
    n=3,4,...:\qquad a_{2n+1}=\alpha \; a_{2n-1} \;- \;
    {(2n-1)(2n-2) \over 4}\; a_{2n-3} \; .\label{3.8b}\end{eqnarray}
%
%
\subsection*{3.3. The basis wave functions. Manifestation of
    vector and spinor space structures}
Having combined the two previous solutions $F_{1}$ and $F_{2}$, we can obtain four
    types of wave functions\footnote{We will change the notation: $F_{1} \Longrightarrow E;
    \; F_{2} \Longrightarrow O$.}
    \begin{equation}\begin{array}{ll} (\mbox{even} \otimes \mbox{even}):\qquad
    &\Phi_{++}=E(a, u^2) \; E(-a, v^2) ,\mm
    (\mbox{odd} \otimes \mbox{odd)}:\qquad&\Phi_{--}=O (a, u) \; O(-a, v\;) \;\; ,\mm
    (\mbox{even} \otimes \mbox{odd)}:\qquad&\Phi_{+-}=E (a, u^2) \; O(-a, v\;) \; ,\mm
    (\mbox{odd} \otimes \mbox{even}):\qquad&\Phi_{-+}=O (a, u)\; E(-a, v^2) \; .
    \end{array}\label{4.1}\end{equation}

Note the behavior of the constructed wave functions:
    \begin{equation}\begin{array}{l}\Phi_{++}(x=0,y=0) \neq 0 \; ,\qquad
    \Phi_{--}(x=0,y=0)=0 \; ,\mm\Phi_{+-}(x>0,y=0)=0 \; ,\qquad \Phi_{-+}(x<0,y=0)=0 \;.
    \end{array}\label{4.3}\end{equation}
Now let us consider which restrictions for the wave functions
$\Psi$ follow from the requirement of single-valuedness. Here two
peculiarities of the parametrization are substantial:
\begin{eqnarray}
\underline{v=0}\;: \; x=+ {u^2 \over 2} \geq 0 , \; y =0 ;
\qquad \underline{u=0}\;: \; x=-{v^2 \over 2} \leq 0 , \; y
=0 \; . \label{4.4}
\end{eqnarray}

\vspace{+3mm}

\unitlength=0.4mm
\begin{picture}(100,50)(-80,0)
\special{em:linewidth 0.4pt} \linethickness{0.6pt}

\put(-50,0){\vector(+1,0){100}} \put(65,-5){$x$}
\put(0,-50){\vector(0,+1){100}} \put(-10, +45){$y$}

\put(0,+1){\line(+1,0){50}} \put(0,+0.5){\line(+1,0){50}}
\put(0,+1.3){\line(+1,0){50}} \put(0,-0.3){\line(+1,0){50}}

\put(+150,0){\vector(+1,0){100}} \put(265,-5){$x$}
\put(200,-50){\vector(0,+1){100}} \put(190, +45){$y$}

\put(2000,+1){\line(-1,0){50}} \put(200,+0.5){\line(-1,0){50}}
\put(200,+1.3){\line(-1,0){50}} \put(200,-0.3){\line(-1,0){50}}

\end{picture}

\vspace{20mm}

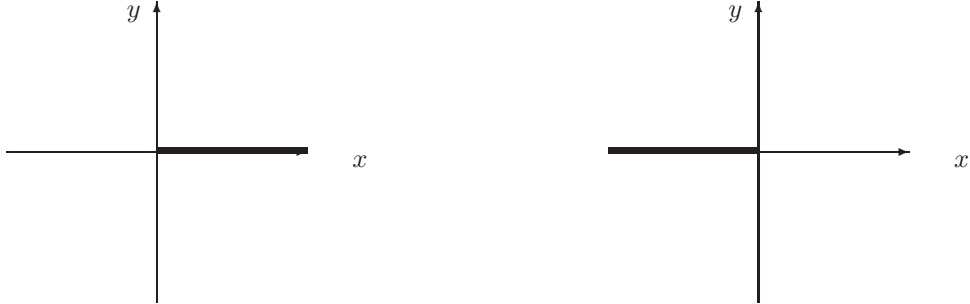
\begin{center}{\small\bf Fig. 12. The peculiarities of the parametrization}\end{center}


Four solutions behave in special regions (see (\ref{4.4})), as follows:
    \begin{eqnarray}&&\Phi_{++} (a;u=0,v)=+ \; \Phi_{++}(a;u=0,-v) \; ,\\
    &&\Phi_{++}(a;+u,v=0)=+ \; \Phi_{++}a;-u,v=0) \; ,\\
    &&\Phi_{--}(a;u =0,+v)=+\; \Phi_{--}(a;u=0,-v)=0\; ,\\
    &&\Phi_{--}(a;u,v=0)=+ \; \Phi_{--}(a;-u,v=0)=0 \; ,\label{4.5}\end{eqnarray}
    \begin{eqnarray}&&\Phi_{+-}(a;u=0,+v)=-\;\Phi_{+-}(a;u=0,-v)\; ,\\
    &&\Phi_{+-} (a;u,v=0)=\Phi_{+-}(a;-u,v=0)=0 \; ,\\
    &&\Phi_{-+}(a;u=0,+ v)= \Phi_{-+}(a;u=0,- v)=0 \; ,\\
    &&\Phi_{-+}(a;+u,v=0)=- \;\Phi_{-+}(a;-u,v=0) \; .\end{eqnarray}
The boundary properties of the constructed wave functions can be illustrated by the following schemes:



\unitlength=0.4mm
\begin{picture}(100,100)(-100,0)
\special{em:linewidth 0.4pt} \linethickness{0.6pt}

\put(-70,+50){$\Phi_{+\;+}$} \put(-50,0){\vector(+1,0){100}}
\put(65,-5){$x$} \put(0,-50){\vector(0,+1){100}} \put(-10,
+45){$y$} \put(+20,+20){\vector(-1,-1){18}}
\put(+24,+20){non-zero} \put(0,0){\circle*{4}}

\put(+100,+50){$\Phi_{-\;-}$} \put(+150,0){\vector(+1,0){100}}
\put(265,-5){$x$} \put(200,-50){\vector(0,+1){100}} \put(+190,
+45){$y$} \put(+220,+20){\vector(-1,-1){18}} \put(+224,+20){zero}
\put(200,0){\circle*{4}}

\put(-70,-80){$\Phi_{+\;-}$} \put(-50,-130){\vector(+1,0){100}}
\put(65,-140){$x$} \put(0,-180){\vector(0,+1){100}} \put(-10,
-90){$y$} \put(+40,-110){\vector(-1,-1){18}} \put(+44,-127){zero}
\put(0,-131){\circle*{4}} \put(0,-131){\line(+1,0){50}}
\put(-50,-128){$+\;+\;+\;+$} \put(-50,-138){$-\;-\;-\;-$}

\put(+100,-80){$\Phi_{-\;+}$} \put(+150,-130){\vector(+1,0){100}}
\put(265,-135){$x$} \put(200,-180){\vector(0,+1){100}} \put(+190,
-90){$y$} \put(+160,-110){\vector(+1,-1){18}}
\put(+140,-120){zero} \put(200,-131){\circle*{4}}
\put(200,-131){\line(-1,0){50}} \put(200,-131.5){\line(-1,0){50}}
\put(+200,-128){$+\;+\;+\;+$} \put(+200,-138){$-\;-\;-\;
- $}

\end{picture}

\vspace{70mm}
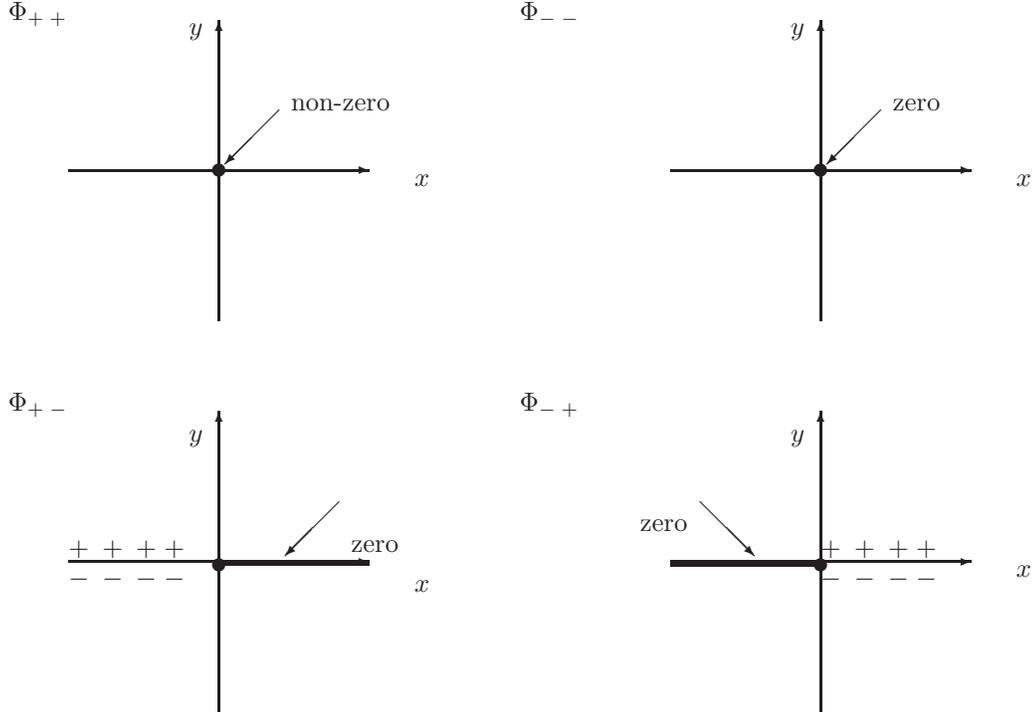
\begin{center}{\small\bf Fig 13. Boundary behavior of the wave functions in the $(x,y)$-plane}\end{center}

So we conclude that the solutions $\Psi$ of the types $(++)$ and $(--)$ are
single-valued in the space with vector structure, whereas the
solutions of the types $(+-)$ and $(-+)$ are not single-valued
in such a space, so these latter types $(+-)$ and $(-+)$ must be discarded.
 However, these solutions ($(+-)$ and $(-+)$) must be retained in the space with
 {\em spinor} structure.

When using the spinor space model, two sets $(u,v)$ and $(-u,-v)$
represent different geometrical points in the spinor space, so the
requirement of single valuedness as applied in the case of spinor
space does not assume that the values of the wave functions
must be equal at the points $(u,v)$ and $(-u,-v)$:
\begin{eqnarray}\Phi (u,v)=\Phi(x,y) \neq\Phi (-u,-v)=\Phi(x',y')\; .\label{4.6d}\end{eqnarray}
The dividing of the basis wave functions into two subsets
may be mathematically formalized with the help of the special discrete
operator acting in the spinor space:
    \begin{eqnarray}\hat{\delta}=\left( \begin{array}{cc} -1&0\\ 0&-1\end{array}\right) ,\qquad
        \hat{\delta}\left(\begin{array}{c} u\\ v\end{array}\right)=\left( \begin{array}{c} -u\\
        -v\end{array}\right) \; .\label{4.6a}\end{eqnarray}
It is easily verified that the solutions which are single-valued in the vector
    space model are eigenfunctions of $\delta$ for the eigenvalue $\delta=+1$:
\begin{eqnarray}\hat{\delta} \; \Phi_{++}(a;u,v)=+ \; \Phi_{++}(a;u,v) \; ,\mm
    \hat{\delta} \; \Phi_{--}(a;u,v)=+ \; \Phi_{--}(a;u,v) \; ,\label{4.6b}\end{eqnarray}
and the additional ones - which are acceptable only in the spinor space
model - are eigenfunctions for the eigenvalue $\delta=-1$:
\begin{eqnarray}\hat{\delta} \; \Phi_{+-}(a;u,v)=- \; \Phi_{+-}(a;u,v) \; ,\mm
    \hat{\delta} \; \Phi_{-+}(a;u,v)=- \; \Phi_{-+}(a;u,v) \; .\label{4.6c}\end{eqnarray}
%
%
\subsection*{{3.4. The form of a diagonalized operator $\hat{A} $ }}
Let us find an explicit form of the operator $\hat{A}$, introduced above
    by the equation $\hat{A} \Psi=a\;\Psi $.\pas
In Cartesian coordinates one has the following representation
\begin{equation}\hat{A}=x \; \left({\partial^2 \over \partial x^2}-{\partial^2 \over \partial y^2}\right)+2y \;
    {\partial^2 \over \partial x \partial y}+{\partial \over \partial x}+x \;
    \left(\;-{\partial^2 \over \partial t^2}+{\partial^2 \over \partial z^2}-m^2\right) \; ;
    \label{5.3a}\end{equation}
which in $(u, v,z)$-coordinates has the form
    \begin{eqnarray}\hat{A}={1 \over 2} \;\left [ \; \left({\partial^2 \over \partial
    u^2}-{\partial^2 \over \partial v^2}\right)-\left({\partial^2\over \partial t^2}-
    {\partial^2 \over \partial z^2}+m^2\right) \; (u^2-v^2) \;\right ] \; .\label{5.3b}\end{eqnarray}
%
%
\subsection*{3.5. Orthogonality and completeness of the bases for vector and spinor spaces }
Now let us consider the scalar multiplication
\begin{eqnarray}\int \Psi^{*}_{\mu'} \; \Psi_{\mu} \; \sqrt{-g} \; dt dz du dv
\; . \label{6.1a}\end{eqnarray}
of the basic constructed wave functions:
\begin{eqnarray}\Psi_{++}(\epsilon, p, a)=e^{i\epsilon t} \; e^{ipz} \;
\Phi_{++}(a;u,v) \; ,\\
\Psi_{--}(\epsilon, p, a)=e^{i\epsilon t} \; e^{ipz} \;
\Phi_{--}(a;u,v) \; ,\\
\Psi_{+-}(\epsilon, p, a)=e^{i\epsilon t} \; e^{ipz} \;
\Phi_{+-}(a;u,v) \; ,\\
\Psi_{-+}(\epsilon, p, a)=e^{i\epsilon t} \; e^{ipz} \;
\Phi_{-+}(a;u,v) \; . \label{6.1b}\end{eqnarray}
where $\mu$ and $\mu'$ stand for generalized quantum numbers.\pas
First of all, we note some interesting integrals\footnote{The arguments ($a;u,v$)
    are omitted here.}:\pas\noindent
\underline{in vector space}
\begin{eqnarray}I_{0}=\int_{0}^{+\infty} dv \int_{-\infty}^{+\infty} du \;
\Phi_{++}^{*} \; \Phi_{--}\; (u^2+v^2) , \label{6.2a}\end{eqnarray}
\underline{in spinor space}
\begin{eqnarray}
I_{1}=\int_{-\infty}^{+\infty} dv \int_{-\infty}^{+\infty}
du \; \Phi_{++}^{*}\; \Phi_{--}\; (u^2+v^2) \; ,\\
I_{2}=\int_{-\infty}^{+\infty} dv \int_{-\infty}^{+\infty}
du \; \Phi_{+-}^{*} \; \Phi_{-+} \; (u^2+v^2) \; ,\\
I_{3}=\int_{-\infty}^{+\infty} dv \int_{-\infty}^{+\infty} du
 \; \Phi_{++}^{*} \; \Phi_{+-} \; (u^2+v^2) \; ,\end{eqnarray}
\begin{eqnarray}
I_{4}=\int_{-\infty}^{+\infty} dv \int_{-\infty}^{+\infty}du
 \; \Phi_{++}^{*} \; \Phi_{-+} \; (u^2+v^2) \; ,\\
I_{5}=\int_{-\infty}^{+\infty} dv \int_{-\infty}^{+\infty}du
 \; \Phi_{--}^{*} \; \Phi_{+-} \; (u^2+v^2) \; ,\\
I_{6}=\int_{-\infty}^{+\infty} dv \int_{-\infty}^{+\infty} du
 \; \Phi_{--}^{*} \; \Phi_{-+} \; (u^2+v^2) \; .\label{6.2b}\end{eqnarray}
All these seven integrals $I_{0}, I_{1}... I_{6}$ are equal to zero,
which means that the constructed functions provide us with an orthogonal basis
for the Hilbert space $\Psi (t,z,u, v)$, where $(u,v,z)$ belong to the
extended (spinor) space model.
%
%
\subsection*{3.6. On matrix elements of physical observables, in vector and spinor spaces}
The question of principle is to determine in which way the transition from vector to
    spinor space model can influence the results of calculation of matrix elements for physical quantities.
    As an example, let us consider matrix elements for operator of coordinates: one may calculate
    the matrix elements of the basic initial coordinates $u, v$ or $x,y$:
\begin{eqnarray} {x }= {u^2-v^2 \over 2}\; , \;
 {y }=uv \;,\qquad \mbox{or}\qquad
 (u,v) . \label{7.1}\end{eqnarray}
Then simple selection rules for the matrix
    elements can be derived\footnote{For simplicity we restrict ourselves only to
    the degeneracy at the discrete quantum number $++,--,+-,-+$, by
    taking $ \epsilon, p, a$ fixed.}:\pas
\underline{in vector space}
\begin{eqnarray}\left. \begin{array}{ccc}
\underline{x_{\mu',\mu}}\qquad&++&--\\[3mm]
++&\neq 0&0\\--&0&\neq 0\end{array}\right.
,\qquad\left. \begin{array}{ccc}
\underline{y_{\mu',\mu}}\qquad&++&--\\[2mm]
++&0&\neq 0\\--&\neq 0& 0\end{array}\right.\end{eqnarray}

\underline{in spinor space}
\begin{eqnarray}
\left. \begin{array}{ccccc}
\underline{x_{\mu',\mu} }\qquad&++&--&+-&-+\\[3mm]
++&\neq 0&0&0&0\\
--&0&\neq 0&0&0\\
+-&0&0&\neq 0&0\\
-+&0&0&0&\neq 0
\end{array}\right.
,\left. \begin{array}{ccccc}
\underline{y_{\mu',\mu}}\qquad&++&--&+-&-+\\[3mm]
++&0&\neq 0&0&0\\
--&\neq 0&0&0&0\\
+-&0&0&0&\neq 0\\
-+&0&0&\neq 0&0
\end{array}\right.
\end{eqnarray}

The same, for the coordinates $u$ and $v$, looks like:\pas

\underline{in vector space}
\begin{eqnarray}
\left. \begin{array}{ccc}
\underline{u_{\mu',\mu}}\qquad&++&--\\[2mm]
++&0&\neq 0\\--&\neq 0&0\end{array}\right. ,
,\qquad\left. \begin{array}{ccc}
\underline{v_{\mu',\mu} }\qquad&++&--\\[2mm]
++&\neq 0&0\\--&0&\neq 0\end{array}\right.\end{eqnarray}

\underline{ in spinor space}
\begin{eqnarray}
\left. \begin{array}{ccccc}
\underline{u_{\mu',\mu} }\qquad&++&--&+-&-+\\[2mm]
++&0&0&0&\neq 0\\
--&0&0&\neq 0&0\\
+-&0&\neq 0&0&0\\
-+&\neq 0&0&0&0
\end{array}\right.
,\left. \begin{array}{ccccc}
\underline{v_{\mu',\mu}}\qquad&++&--&+-&-+\\[2mm]
++&0&0&\neq 0&0\\
--&0&0&0&\neq 0\\
+-&\neq 0&0&0&0\\
-+&0&\neq 0&0&0
\end{array}\right.
\end{eqnarray}
%
%
\subsection*{3.7. Schr\"{o}dinger equation}
The study of the analytical properties of the Klein-Fock-Gordon wave solutions
    in vector and spinor space models is still applicable, with slight changes, to
    the non-relativistic Schr{\"{o}}dinger equation as well:
\begin{eqnarray}i \hbar {\partial \over \partial t} \Psi=- {\hbar^2 \over 2m}\;\left [\;
    {\partial^2 \over \partial z^2}+{1 \over u^2+v^2 }\; \left({\partial^2 \over \partial u^2}+
    {\partial^2 \over \partial v^2}\right)\;\right ] \Psi \; ,\label{8.1a}\end{eqnarray}
where the substitution for the wave functions is the same
\begin{eqnarray}\Psi (t,u,v,z)=e^{-i\epsilon t /\hbar} \; e^{ipz / \hbar} \;U(u) V(v) \; ,\end{eqnarray}
and then, the equation for $U(u)V(v)$ is
\begin{eqnarray}\left [ \; {\hbar^2 \over 2m}\; \left({\partial^2 \over \partial u^2}+
    {\partial^2 \over \partial v^2}\right)\; \;+\;\left(\epsilon -{p^2 \over 2m}\right)
    (u^2+v^2))\;\right ] U(u) V(v)=0 \; .\label{8.1b}\end{eqnarray}
%
%
\subsection*{3.8. Conclusions to Section 3}
We shall further infer several quantum mechanical consequences while
    changing the vector geometrical model of the physical space to the spinor one.\pas
The extension procedure is performed in cylindrical parabolic coordinates, $G(t,u,v,z) \Longrightarrow
    \tilde{G}(t,u,v,z)$. This is done through expansion
    of the region  $G$, so that instead of the half plane $(u,v>0)$ now the entire plane $(u,v)$ should be used,
    accompanied with new identification rules for the boundary points. In the Cartesian picture, this procedure
    corresponds to taking the two-sheet surface $(x',y') \oplus (x'',y'')$ instead of the one-sheet surface
    $(x,y)$.\pas
The solutions of the Klein--Fock--Gordon and Schr\"{o}dinger equations
$ \Psi_{\epsilon, p, \;a}=e^{i\epsilon t} e^{ipz} U_{a}(u)
V_{a}(v) $ are constructed in terms of parabolic cylindric functions\footnote{We denoted the
    separating constant by $a$.}. Given the quantum numbers $\epsilon, p, a$, four types of solutions are
    possible: $\Psi_{++}, \Psi_{--}; $ $\; \Psi_{+-},\Psi_{-+}$.\pas
The first two ones, $\Psi_{++}$ and $\Psi_{--}$, provide us with single-valued functions of the vector
    space points, whereas the last two, $\Psi_{+-}$ and $\Psi_{-+}$, have discontinuities in the frame of
    vector spaces, and therefore they must be discarded in this model.
    All the four types of functions are continuous ones while regarded in the spinor space.\pas
It is established that all solutions $\Psi_{++}, \Psi_{--}, \; \Psi_{+-}$ and $\Psi_{-+}$,
    are orthogonal to each other, provided that integration is done over the extended
    region of integration which covers (corresponds to)  the spinor space.

    \pas
Some simple selection rules for matrix elements of the vector and spinor coordinates,
    $(x,y)$ and $(u,v)$, respectively, are further derived. The selection rules for
    $(u,v)$ are substantially different in vector spaces compared to spinor spaces.
%
%
\section{Some relevant topics}

The problem we  addressed in the present paper  can be relevant to a number of other  topics:
 the relation between
    the Dirac--Schwinger quantization rule and the superposition principle
    in quantum mechanics; the manifestation of spinor space structure in
    classifying the solutions of the Dirac equation and for the matrix elements
    which are related to physical quantities; spinors in polarization optics; the Jones
    formalism for completely and partly polarized light; General Relativity and
    Riemannian space-time models with spinor structure and tetrad (vierbein) formalism.

%
\section*{Acknowledgment}
The present work was developed under the auspices of Grant 1196/2012 - BRFFR-RA No.
    F12RA-002, within the cooperation framework between Romanian Academy
    and Belarusian Republican Foundation for Fundamental Research.\par
The authors wish to thank to the organizers of the joint event {\em Colloquium on Differential Geometry},
    and {\em The IX-th International Conference on Finsler Extensions of
    Relativity Theory (FERT 2013)}, held between 26 -- 30 August 2013 in Debrecen, Hungary.
    for their worm hospitality.
Also, V. Red'kov, O. Veko and V. Balan are thankful to Prof. D. Pavlov for the support
    provided for the participation in the event {\em FERT 2013}.
\noindent
Elena Ovsiyuk, Olga Veko\\Mozyr State Pedagogical University, Belarus.\\[1mm]
Alexandru Oan\u{a}, Mircea Neagu\\University Transilvania of Bra\c{s}ov, Romania.\\[1mm]
Vladimir Balan\\University Politehnica of Bucharest, Romania.\\[1mm]
Victor Red'kov\\B.I. Stepanov Institute of Physics, NAS of Belarus, Minsk, Belarus.
\end{document}